\documentclass[twocolumn,floatfix]{revtex4}
\pdfoutput=1
\usepackage{graphicx}
\usepackage{dcolumn}
\usepackage{longtable}
\usepackage{amsmath}
\usepackage{amssymb}
\usepackage{color}
\usepackage{isotope}

\begin{document}
\title{Preponderance of triaxial shapes in atomic nuclei predicted by the proxy-SU(3) symmetry}

\author
{Dennis Bonatsos$^1$(corresponding author), Andriana Martinou$^1$,  S.K. Peroulis$^1$, D. Petrellis$^2$, P. Vasileiou$^3$, T.J. Mertzimekis$^3$, and N. Minkov$^4$ }

\affiliation
{$^1$Institute of Nuclear and Particle Physics, National Centre for Scientific Research ``Demokritos'', GR-15310 Aghia Paraskevi, Attiki, Greece}

\affiliation
{$^2$ Nuclear Physics Institute, Czech Academy of Sciences, CZ-250 68 \v{R}e\v{z} near Prague, Czech Republic}

\affiliation
{$^3$  Department of Physics, National and Kapodistrian University of Athens, Zografou Campus, GR-15784 Athens, Greece}

\affiliation
{$^4$Institute of Nuclear Research and Nuclear Energy, Bulgarian Academy of Sciences, 72 Tzarigrad Road, 1784 Sofia, Bulgaria}

\begin{abstract}

The proxy-SU(3) symmetry predicts, in a parameter-free way, based only on the Pauli principle and the short-range nature of the nucleon-nucleon interaction, non-vanishing values of the collective variable $\gamma$ almost everywhere across the nuclear chart. Substantial triaxiality with $\gamma$ between 15$^{\rm o}$ and 45$^{\rm o}$ is proved to be expected along horizontal and vertical stripes on the nuclear chart, covering the nucleon numbers 22-26, 34-48, 74-80, 116-124, 172-182. Empirical support for these stripes is found by collecting all even-even nuclei for which the first two excited $2^+$ states are known, along with the $B(E2)$s connecting them, as well as the second $2^+$ state to the ground state. The stripes are related to regions in which oblate SU(3) irreducible representations appear, bearing similarity to the appearance of triaxiality within the SU(3)$^*$ dynamical symmetry of the interacting boson model-2. Detailed comparisons of the proxy-SU(3) predictions to the data and to predictions by state-of-the-art Monte Carlo shell model calculations for deformed $N=94$, 96, 98 isotones in the rare earth region show good overall agreement, with the exception of $Z=70$ and $N=94$, which correspond to fully symmetric proxy-SU(3) irreps, suggesting that the latter are an artifact of the method which can be amended by considering the influence of the neighboring irreps. 
 
\end{abstract}

\maketitle

E-mail: bonat@inp.demokritos.gr

\section{Introduction}

Triaxial nuclei, i.e., nuclei with three unequal axes, have a long history in nuclear structure studies. Soon after the introduction of the collective model of Bohr and Mottelson \cite{Bohr1952,Bohr1998b} in 1952, in which the nuclear properties are described in terms of the collective variables $\beta$ and $\gamma$, indicating the departure from sphericity and from axial symmetry respectively, the rigid triaxial rotor model has been introduced by Davydov {\it et al.} \cite{Davydov1958,Davydov1959} in 1958, in 
which $\gamma$ assumes a constant (rigid) value. The model has been solved analytically for maximal triaxiality ($\gamma=30^{\rm o}$) in 1975 by Meyer-ter-Vehn 
\cite{MeyerterVehn1975}. In the algebraic framework of the interacting boson model (IBM), introduced by Arima and Iachello \cite{Arima1975,Iachello1987} in 1975, the nuclear  collective properties are described in terms of $s$ and $d$ bosons, bearing angular momentum 0 and 2, respectively and corresponding to correlated pairs of valence nucleons, giving rise to three dynamical symmetries, U(5), SU(3), and O(6), corresponding to spherical, axially deformed, and $\gamma$-unstable (soft towards triaxial deformation) nuclei,  respectively. Triaxial shapes have been introduced \cite{Dieperink1982} in 1982 in the framework of IBM-2 \cite{Iachello1987}, in which distinction is made between protons and neutrons. In the IBM-2 framework the SU(3)$^*$ dynamical symmetry \cite{Dieperink1982,Walet1987} appears, in which protons correspond to valence particles and neutrons correspond to valence holes, or {\it vice versa}. 

In the realm of mean field models, early studies of triaxial nuclei include Hartree--Fock (HF) calculations in 1969 \cite{Giraud1969}, Hartree--Bogolyubov (HB) calculations in 1978 \cite{Girod1978}, as well as Hartree--Fock--Bogolyubov (HFB) calculations in 1983 \cite{Girod1983}, employing the Gogny D1 interaction. The importance of employing angular momentum projection in order to single out minima with triaxial shapes in the energy surfaces in the HFB framework has been realized in 1984 \cite{Hayashi1984}. 
Angular momentum projection has also been used, along with configuration mixing, in relativistic mean field calculations \cite{Yao2010} for triaxial nuclei.

The spherical shell model, introduced by Mayer and Jensen \cite{Mayer1949,Haxel1949,Mayer1955} in 1949, based on the isotropic 3-dimensional harmonic oscillator (3DHO) plus a spin-orbit interaction, as well as the deformed shell model, introduced by Nilsson \cite{Nilsson1955,Nilsson1995} in 1955, based on an axially deformed 3DHO, cannot readily accommodate triaxial shapes. This has been achieved by Sheikh and Hara \cite{Sheikh1999} in 1999 with the introduction of the triaxial projected shell model (TPSM), in which triaxiality is explicitly taken into account and 3-dimensional angular momentum projection is performed from the triaxial Nilsson+BCS wave function \cite{Sheikh1999}. Since the computational effort needed in the TPSM is negligible, in comparison to the full shell model approach, detailed calculations for several nuclei have been performed in this framework \cite{Rouoof2024}. A still highly demanding, but much improved, in comparison to the conventional shell model, computational path has been suggested by Otsuka and collaborators \cite{Honma1995}, by introducing the Monte Carlo shell model (MCSM) methodology. Detailed calculations for some heavy deformed nuclei by the MCSM method have been recently achieved \cite{Tsunoda2021,Otsuka2024}.   

A different microscopic path towards nuclear deformation has been discovered in 1958 by Elliott \cite{Elliott1958a,Elliott1958b,Elliott1963}, who proved that quadrupole deformation can occur within the spherical shell model, revealed through the use of the SU(3) symmetry underlying the nuclear 3DHO shells \cite{Wybourne1974}. The spin-orbit interaction is known to break the SU(3) symmetry beyond the $sd$ nuclear shell \cite{Mayer1949,Haxel1949,Mayer1955}, but its restoration in heavier shells has been suggested by different approximate ways, including the pseudo-SU(3) (1973) \cite{RatnaRaju1973,Draayer1983,Draayer1984}, quasi-SU(3) (1995) \cite{Zuker1995,Zuker2015}, and proxy-SU(3) (2017) \cite{Bonatsos2017a,Bonatsos2017b,Bonatsos2023b} schemes, in which triaxiality is readily obtained.   

A bridge between the collective model of Bohr and Mottelson and the algebraic approach using SU(3) has been achieved in 1988 \cite{Castanos1988}, by mapping the invariants of the collective model onto the invariants of the rigid rotor (the Casimir operators of SU(3)). Through this mapping the collective variables $\beta$ and  $\gamma$ of the collective model are connected to the Elliott quantum numbers $\lambda$ and $\mu$, characterizing the SU(3) irreducible representations (irreps) $(\lambda,\mu)$ in the Elliott notation \cite{Elliott1958a}. For the $\gamma$ variable the connection reads \cite{Castanos1988}
\begin{equation}\label{mu}
\gamma = \arctan \left( {\sqrt{3} (\mu+1) \over 2\lambda+\mu+3}  \right).
\end{equation}
The variable $\gamma$ is known to assume values between 0 and 60$^{\rm o}$ \cite{Bohr1952,Bohr1998b}, corresponding to prolate (rugby ball like) and oblate (pancake like) axially symmetric deformed shapes, with triaxial shapes occurring in between and maximal triaxiality seen at $30^{\rm o}$. Purely prolate (oblate) SU(3) irreps have the form $(\lambda,0)$ 
[$(0,\mu)$], while $(\lambda,\mu)$ irreps are in general triaxial. Prolate-like (oblate-like) irreps with $\lambda>\mu$ ($\lambda<\mu$) are usually referred to simply as prolate (oblate), while irreps with $\lambda=\mu$ correspond to the maximum triaxiality. Systems of bosons obviously correspond to the fully symmetric  $(\lambda,0)$ irreps
\cite{Iachello1987}, while for systems of boson holes the irreps [$(0,\mu)$] have been used \cite{Dieperink1982,Walet1987}.

In recent years, numerous theoretical studies have been performed not only for isolated triaxial nuclei, but also in some series of isotopes in which triaxial deformation is expected, including TPSM calculations \cite{Bhat2014,Zhang2015b,Jehangir2021,Rajput2022}, as well as mean-field calculations extending from HFB \cite{Hayashi1984,Heenen1993,Oi2003,RodriguezGuzman2010,RodriguezGuzman2010b,Chen2017} to relativistic mean field (RMF) \cite{Yao2010,Niksic2010,Xiang2016} and covariant density functional theory (CDFT) \cite{Abusara2017a,Yang2021a,Nomura2021a,Nomura2021b,ElBassem2024}. However, it seems that an effort to locate regions of the nuclear chart in which triaxiality would be preferable is lacking. 

In the present manuscript we address the above question using the proxy-SU(3) approximation to the shell model, 
in a way similar to the prediction of regions of the nuclear chart in which shape coexistence can be expected to occur \cite{Martinou2021a,Martinou2023,Bonatsos2023}. In Section II the relevant proxy-SU(3) predictions are given, to be compared with existing empirical information in Section III and in Section IV to recent numerical results for some specific deformed nuclei, recently obtained through state-of-the-art configuration interaction calculations in the shell model framework \cite{Tsunoda2021,Otsuka2024}.

\section{Proxy-SU(3) predictions} \label{proxy}

The SU(3) symmetry of the isotropic 3DHO is destroyed beyond the $sd$ shell because of the spin-orbit interaction, which within each major shell pushes a bunch of orbitals (the deserting orbitals) to the shell below \cite{Mayer1949,Haxel1949,Mayer1955}. As a result, each major shell remains with the rest of its original orbitals, plus the ones coming down from the shell above (the intruder orbitals). Within the proxy-SU(3) scheme \cite{Bonatsos2017a,Bonatsos2017b,Bonatsos2023b}, the SU(3) symmetry is restored by using the deserting orbitals as "proxies" of the intruder orbitals (except the intruder orbital with the highest projection of the total angular momentum, which however lies at the top of the shell and thus is empty for most nuclei). 

Within the proxy-SU(3) scheme \cite{Bonatsos2017a,Bonatsos2017b,Bonatsos2023b} each nucleus is characterized by an SU(3) irrep $(\lambda,\mu)$, which is the most stretched irrep $(\lambda_p+\lambda_n, \mu_p+\mu_n)$ coming from the highest weight irrep \cite{Martinou2021b} corresponding to the valence protons $(\lambda_p, \mu_p)$ and the highest weight irrep coming from the valence neutrons $(\lambda_n, \mu_n)$.
It has been shown \cite{Martinou2021b} that the highest weight irrep is the most symmetric irrep allowed by the restrictions imposed by the Pauli principle and the short range nature of the nucleon-nucleon interaction. 

The necessary SU(3) irreps for the valence protons or neutrons are easily obtained from Table I of Ref. \cite{Bonatsos2017b}, with several examples appearing in Tables II and III of the same reference. Additional tables are given in Ref. \cite{Sarantopoulou2017}. From Table I of Ref. \cite{Bonatsos2017b} and Table 1 of Ref. \cite{Sarantopoulou2017} we see that within the major nuclear shells, created by the spin-orbit interaction and bordered by the magic numbers 28, 50, 82, 126, 184  
\cite{Martinou2020},  SU(3) irreps with $\lambda \leq \mu$ (oblate-like or maximally triaxial) occur for the nucleon numbers 22-26, 42-48, 74-80, 116-124, 172-182. For example, the major shell 50-82 corresponds to the U(15) column in Table I of Ref. \cite{Bonatsos2017b}, in which irreps with $\lambda\leq \mu$ appear for 24-30 valence particles, i.e. for 74-80 particles in the 50-82 shell. Furthermore, as discussed extensively in Ref. \cite{Martinou2020}, in addition to the shells created by the spin-orbit interaction, the pure isotropic 3D-HO shells, bordered by the magic numbers  2, 8, 20, 40, 70, 112, 168, may play a distinct role in some cases, as for example in the dual-shell mechanism \cite{Martinou2021a} revealing the regions of the nuclear chart in which shape coexistence may be expected to occur (see Table 7 and Fig. 1 of Ref. \cite{Martinou2020} for further details on the connection of the proxy-SU(3) scheme to the spherical shell model through a unitary transformation).  For the present purpose of determining regions of the nuclear chart in which nuclei with increased triaxiality may occur, only the sometimes strong subshell closure at 40 nucleons \cite{Sorlin2008} is expected to play a role. In the same way as above, we see that within the 20-40 shell, bearing the U(10) symmetry, irreps with $\lambda\leq \mu$ occur for 14-20 valence nucleons, i.e., for 34-40 particles.

In view of the above, nuclei with substantial triaxiality are expected to occur within horizontal and vertical stripes in the nuclear chart, bordered by the nucleon (proton or neutron) numbers 22-26, 34-48, 74-80, 116-124, 172-182, shown in Fig. 1. This will happen because for these nuclei at least one of the involved h.w. irreps (protons or neutrons) will be oblate-like, thus carrying on a substantial value of  $\mu$ to the total irrep $(\lambda,\mu)$ characterizing the properties of the nucleus. It should be pointed out that these predictions will be completely parameter-free, based only on the proxy-SU(3) symmetry and the consequences of the Pauli principle and the short-range nature of the nucleon-nucleon interaction, which favor the h.w. irreps \cite{Martinou2021b}. 

In Fig. 1 the value of $\gamma$ obtained from Eq. (\ref{mu}) for all experimentally known \cite{ensdf} nuclei with $Z=18$-80 and $N=18$-124 is shown. We see that very few nuclei exhibit $\gamma$ close to 0 (pure prolate shape) or 60$^{\rm o}$ (pure oblate shape), while the great majority of nuclei shows $\gamma$ between 5$^{\rm o}$ and 55$^{\rm o}$. This result is in qualitative agreement with recent studies \cite{Grosse2022} based on the analysis of data for several observables in heavy nuclei, showing extensive presence of triaxiality in them.  

In Fig. 2 we keep only the nuclei exhibiting substantial triaxiality, with $\gamma$ between 15$^{\rm o}$ and 45$^{\rm o}$. We see that these tend to be aligned along the regions (indicated in Fig. 2 by green stripes) in which oblate SU(3) irreps for the valence protons or neutrons are predicted by the proxy-SU(3) symmetry, thus resulting in substantial $\mu$ values in the 
$(\lambda,\mu)$ SU(3) irreps, resulting in larger $\gamma$ values. The few nuclei not falling within the green stripes in Fig. 2 are indeed touching them. It should be noticed that the choice of nuclei with $\gamma$ between 15$^{\rm o}$ and 45$^{\rm o}$ is arbitrary, having the meaning of considering nuclei in the region of $\pm 15^{\rm o}$ around the value of maximal triaxiality, occurring at 30$^{\rm o}$.     

The findings of Fig. 2 are in qualitative agreement with the basic idea underlying the SU(3)$^*$ symmetry of IBM-2 \cite{Dieperink1982,Walet1987}, in which either the valence protons or the valence  neutrons  should be holes, which correspond to the upper half of the relevant shell. For example, in a nucleus with $N_p$ pairs of valence proton particles and $N_n$ pairs of valence neutron holes, the corresponding SU(3) irreps are $(2N_p,0)$ and $(0,2N_n)$, giving the total irrep $(2N_p,2N_n)$, which will present substantial triaxiality. 

It should be pointed out that the values of the collective variables $\beta$ and $\gamma$ obtained \cite{Bonatsos2020a} in the framework of the  pseudo-SU(3) scheme 
\cite{RatnaRaju1973,Draayer1983,Draayer1984} using the h.w. irreps,  have been found to be quite close to the predictions of the proxy-SU(3) scheme, despite the fact that the approximations made and the unitary transformations used in the two schemes are very different. In addition to adding credibility to both approximation schemes, this result suggests that a similar study carried out in the pseudo-SU(3) framework would produce results very similar to these shown in Figs. 1 and 2.   

\section{Comparison to empirical systematics}\label{emp} 

It would be interesting to examine to which extent these predictions are supported by the data. In this direction one can use the analytical expressions derived within the Davydov model \cite{Davydov1958,Davydov1959}. One can obtain $\gamma$ directly from the energy ratio 
\begin{equation} \label{R}
R= {E(2_2^+) \over E(2_1^+)} 
\end{equation}
through the expression \cite{Casten2000}
\begin{equation} \label{g}
\gamma = {1\over 3} \sin^{-1} \left( {3\over R+1} \sqrt{R\over 2}  \right). 
\end{equation}
One can also obtain $\gamma$ from the ratio of transition rates (branching ratio) 
\begin{equation} \label{R2}
R_2= {B(E2; 2_2^+ \to 2_1^+)  \over B(E2; 2_2^+ \to 0_1^+)} 
\end{equation}
through the expression \cite{Casten2000}
\begin{equation}
R_2 = {20\over 7} { {\sin^2 3\gamma  \over 9-8 \sin^2 3\gamma} \over   1-{3-2\sin^2 3\gamma \over \sqrt{9-8\sin^2 3\gamma} }  } . 
\end{equation} 

The ratios $R$ and $R_2$ are shown in Fig. 3. It is clear that $R$ starts from the value 2 at $\gamma=30^{\rm o}$ and raises towards infinity for $\gamma=0$, while $R_2$ shows the opposite behavior, starting from 1.43 at $\gamma=0$ and raising towards infinity at  $\gamma=30^{\rm o}$. At $\gamma=15^{\rm o}$ one has $R=6.85$ and $R_2=2.71$. 

In Fig. 4 all nuclei with known \cite{ensdf} $2^+_1$ and $2_2^+$ levels are shown, divided into two groups, these with $R<6.85$ (expected to have $\gamma > 15^{\rm o}$) and those with $R>6.85$ (expected to have $\gamma < 15^{\rm o}$). We see that many nuclei belong to the first group. However, not all of them are expected to exhibit strong triaxiality, since $R=2$ also corresponds to the simple vibrator with U(5) symmetry \cite{Iachello1987}. 

In Table I all nuclei with experimentally known $2^+_1$ and $2_2^+$ levels, as well as $B(E2; 2^+_2\to 0^+_1)$ and $B(E2; 2^+_2\to 2^+_1)$ transition rates, are shown. 
The experimental $R_{4/2}=E(4_1^+)/E(2_1^+)$ ratio, a well-known \cite{Casten2000} indicator of collectivity, is also shown, in order to facilitate the recognition of nuclei being close to the simple vibrator value of $R_{4/2}=2$, mentioned above. 
 
In Fig. 5 the nuclei of Table I are subdivided into two groups, these with $R<6.85$ and $R_2>2.71$ (expected to have $\gamma > 15^{\rm o}$) and those with $R>6.85$ and $R_2<2.71$ (expected to have $\gamma < 15^{\rm o}$). 
 We see that by taking into account the value of the branching ratio, the number of candidates for substantial triaxiality is drastically reduced, while most of the remaining candidates are aligned along the regions (indicated in Fig. 5 by green stripes) for which oblate SU(3) irreps are predicted by the proxy-SU(3) symmetry, in  agreement with the situation appearing in Fig. 2. A substantial deviation is seen in the $Z=50$-56, $N=62$-72 region, in which nine nuclei with $\gamma   > 15^{\rm o}$ are lying outside the green stripes.

\section{Comparison to shell model predictions} \label{MCSM}

In a recent series of papers by Otsuka and collaborators \cite{Otsuka2019,Tsunoda2021,Otsuka2024} the appearance of triaxiality in heavy deformed nuclei like 
\isotope[166][68]{Er}, traditionally considered as prolate rotors with a $\gamma$-band appearing close to the ground state band, have been considered in the framework of state-of-the-art Monte Carlo shell model (MCSM) \cite{Honma1995} calculations. It has been argued \cite{Otsuka2024} that a relatively small amount of basic modest triaxiality coming from symmetry restoration appears in all nuclei, but in addition a large amount of prominent triaxiality due to the pn tensor monopole interaction and the pn central multipole (mainly hexadecapole) interaction also appears. Detailed results of highly demanding MCSM computations have been reported \cite{Otsuka2024} for some deformed $N=94$, 96, 98 isotones in the rare earth region. 

It is of interest to see to what extend the bare-bones, parameter-free predictions of the proxy-SU(3) scheme, which are obtained practically at no computational cost, agree with the results of MCSM highly sophisticated and time-consuming calculations. 

In the left column of Fig. 6 the proxy-SU(3) and MCSM predictions for $\gamma$ for the $N=94$, 96, 98 series of isotones are compared to the experimental values extracted from the ratios $R$ and $R_2$ as described in Sec. \ref{emp} and illustrated in Fig. 3. The following observations can be made. 

a) In all cases there is reasonable agreement between the empirical $R$ and $R_2$ values, with $R$ being higher than $R_2$, either slightly, as for $N=94$, or more explicitly, as for $N=96$, 98.

b) In all cases the MCSM predictions for the $K=2$  band are higher than the predictions for the ground state ($K=0$) band.  

c) The $N=96$, 98 cases are quite similar, with the proxy-SU(3) predictions (except for $Z=70$, to be discussed below) being close to the empirical $R$ values and the MCSM predictions being close to the empirical $R_2$ values, with the former pair lying considerably higher than the latter. In other words, there is a higher group, consisting of 
proxy-SU(3) and $R$, and a lower group, consisting of MCSM and $R_2$. 

d) The $N=94$ case presents some modifications. The higher group consists of the $R$ and $R_2$ empirical values and the MCSM predictions for the $K=2$ bands, while the lower group is made of the proxy-SU(3) predictions and the MCSM predictions for the $K=0$ band. In other words, the proxy-SU(3) predictions have moved from the top group to the bottom group, while the prediction for $Z=70$ still remains highly problematic as before. 

These observations can be summarized as follows.

1) The MCSM predictions are in general close to the empirical $R_2$ values, with the exception of the predictions for the ground state band at $N=94$, which are somewhat lower.   

2) The proxy-SU(3) predictions are close to the empirical $R$ values, except for $N=94$, in which they are lower. 

3) The proxy-SU(3) predictions for $Z=70$ are too low in all cases. 

The above observations reveal that there is a problem of underestimation of the $\gamma$ value in proxy-SU(3) for $Z=70$ and $N=94$. This problem appears  to be due to the fact that the relevant h.w. irreps, which are (20,0) and (36,0) respectively, as one can see in Table I of Ref. \cite{Bonatsos2017a}, have $\mu=0$, thus dropping the value of $\gamma$ obtained from Eq. (\ref{mu}). It seems that the appearance of hw irreps with $\mu=0$ is a rather extreme limiting case in the proxy-SU(3) framework, which should be avoided by considering the contributions of the next higher weight irreps, which have $\mu\geq 4$, since no irreps with $\mu=2$ occur
among the stretched irreps corresponding to nuclei, as seen in Refs. \cite{Bonatsos2017b,Sarantopoulou2017}. Preliminary calculations involving the next higher weight (nhw) irrep along with the hw irrep, described in Appendices A and B, seem to suggest that the modifications occurring for the predictions of $\gamma$ are negligible if the h.w. irrep has $\mu\neq 0$ (which means $\mu \geq 4$), but they are sizeable if the h.w. irrep has $\mu=0$. 

The $\mu=0$ problem appears most severely at \isotope[164][70]{Yb}$_{94}$, in which both the proton and neutron h.w. irreps have $\mu=0$, resulting in the total irrep (56,0). This creates a major problem, since the lowest $K=2$ band (traditionally called the $\gamma$-band) and the lowest $K=4$ band (traditionally called the $\gamma\gamma$ band) should be accommodated in the same framework, as suggested by the MCSM results \cite{Otsuka2024}. In most nuclei the proxy-SU(3) scheme predicts h.w. irreps with $\mu \geq 4$ \cite{Bonatsos2017b,Sarantopoulou2017}, thus the $K=0$ ground state band and the lowest $K=2$ and $K=4$ bands belong to the same SU(3) irrep, so that interband transitions are allowed to occur within the symmetry. The same situation occurs \cite{Bonatsos2020a} within the pseudo-SU(3) scheme \cite{RatnaRaju1973,Draayer1983,Draayer1984}, although it is based on different approximations. It appears that the common framework for the lowest $K=0$, 2, 4 bands is a requirement of the Pauli principle and the short-range nature of the nucleon-nucleon interaction within the proxy-SU(3) and pseudo-SU(3) schemes, manifested in the MCSM framework by the pn tensor monopole interaction and the pn central multipole interaction \cite{Otsuka2024}. It seems that this requirement appears in all models using fermions, irrespectively of  the use or not of algebraic techniques. This requirement contradicts the simple picture of the SU(3) limit of IBM-1 \cite{Iachello1987}, in which for a nucleus corresponding to $N$ bosons the ground state band is sitting alone in the $(2N,0)$ irrep, the lowest $K=2$ band sits in the $(2N-4,2)$ irrep, and the lowest $K=4$ band sits in the $(2N-8,4)$ irrep, with interband transitions among them being forbidden within the SU(3) symmetry, since they belong to different irreps, thus leading to the need to break the SU(3) symmetry in order to allow them to occur. 

The results of the schematic calculation described in Appendix B are shown in the right column of Fig. 6, in which the following observations can be made. 
      
1) The MCSM predictions are in general close to the empirical $R_2$ values, with the exception of the predictions for the ground state band at $N=94$, which are somewhat lower.   

2) The proxy-SU(3) predictions are close to the empirical $R$ values. 

3) The sudden drops of the proxy-SU(3) predictions at $Z=70$ have been eliminated. 

It can therefore be concluded that the MCSM calculations provide guidance for the amendment of the proxy-SU(3) scheme in the cases in which irreps with $\mu=0$ appear. 

\section{Conclusions}

The main conclusions of the present work are summarized here.

a) The proxy-SU(3) symmetry predicts some degree of triaxiality almost all over the nuclear chart, with substantial values of triaxiality occurring within horizontal and vertical stripes on the nuclear chart covering the nucleon numbers 22-26, 34-48, 74-80, 116-124, and 172-182. These predictions, derived with minimal numerical effort, are completely parameter-free and are based on the Pauli principle and the short-range nature of the nucleon-nucleon interaction. 

b) The stripes predicted by the proxy-SU(3) symmetry are supported by the data, by singling out nuclei of which the experimental energy ratio $R$ and the experimental $B(E2)$s ratio $R_2$ correspond to $\gamma$ values above 15$^{\rm o}$, as extracted through the formalism of the Davydov model.  

c) The comparison between the proxy-SU(3) predictions and the values predicted by state-of-the-art, computationally demanding MCSM calculations, for the collective parameter $\gamma$ for the $N=94$, 96, 98 series of isotones reveal large deviations occurring at $Z=70$ and $N=94$, which suggest that h.w. SU(3) irreps in proxy-SU(3) with $\mu=0$ (fully symmetric irreps) are an extreme limit of the method used, which should be avoided  by taking into account the influence of the next higher weight (nhw) SU(3) irreps. After taking the nhw irreps into account, the proxy-SU(3) predictions are close to these corresponding to the experimental energy ratio $R$, while the MCSM predictions are close to these corresponding to the experimental $B(E2)$ ratio $R_2$, the former couple of predictions being systematically higher than the latter.

d) The stripes predicted by proxy-SU(3) are compatible with the basic idea underlying the SU(3)$^*$ symmetry of IBM-2, namely that triaxial shapes appear when one type of the valence nucleons (protons or neutrons) corresponds to particles and the other one to holes.   

Predominance of triaxial shapes in some superheavy nuclei (Flerovium ($Z=114$) isotopes) has been recently predicted by triaxial symmetry-conserving configuration-mixing (SCCM) calculations \cite{Egido2020}. The extension of the proxy-SU(3) predictions to superheavy nuclei is a straightforward task. 

What has been ignored in the present study is the distinction between rigid and soft triaxiality, which is usually made by looking at the odd-even staggering of the levels of the $K=2$ band \cite{Zamfir1991,McCutchan2007}. Based on the available data \cite{ensdf}, very few nuclei have been found \cite{McCutchan2007} to correspond to rigid triaxiality. Efforts to find signatures distinguishing rigid from soft triaxiality based on the levels of both the ground state and $K=2$ bands \cite{Casten2020,Bonatsos2021} are in progress.  

\section*{Acknowledgements} 

Support by the Bulgarian National Science Fund (BNSF) under Contract No. KP-06-N48/1  is gratefully acknowledged.

\section*{Appendix A: Highest weight irreducible representations of SU(3) and their next ones}

The highest weight (hw) irreducible representations of SU(3) are the most symmetric irreps for a given number of nucleons allowed by the Pauli principle and the short-range nature of the nucleon-nucleon interaction \cite{Martinou2021b}. The hw irreps for U(15) and U(21) (the symmetries corresponding to the valence protons and neutrons, respectively, in the heavy rare earth region within the proxy-SU(3) scheme) are shown in Table II, along with the next higher weight irreps, which are the second preference of the Pauli principle and the sort-range nature of the nucleon-nucleon interaction, calculated through the code UNTOU3 \cite{Draayer1989}. The hw irreps for U(6), U(10), and U(28) are also shown for comparison. The following observations can be made. 

a) If the hw irrep is $(\lambda,\mu)$, the next higher weight irrep is always $(\lambda+2, \mu-4)$, except in the cases in which $\mu=0$, 2, in which this is obviously impossible. 

b) $\mu=2$ occurs only for particle number $M=4$, while $\mu=0$ occurs for $M=2$, 6, 12, 20, 30, 42. 

This peculiarity has certain physical consequences. The hw irreps with $\mu \geq 4$ can accommodate the ground state band, which has $K=0$, as well as the lowest $K=2$ band and the lowest $K=4$ band. Since these three bands belong to the same irrep, they can be connected through electromagnetic transitions without having to break the SU(3) symmetry. 

In contrast, the hw irreps with $\mu=0$ can accommodate only the ground state band (gsb), while the $K=2$ and $K=4$ bands will have to be accommodated within a higher irrep, thus no electromagnetic transitions connecting them to the gsb would be allowed. It becomes clear that in these cases both the hw irrep and the next higher weight (nhw) irrep should be simultaneously taken into account in order to describe the $K=0$, 2, 4 bands within the same framework. 

As a simple approximation, we assume a 50\% contribution from each of the two irreps.  

\section*{Appendix B: Application to the $N=94$, 96, 98 series of isotones} 

In Table III the nuclei affected by the replacement of the hw irrep by the next higher weight (nhw) irrep in the $N=94$, 96, 98 series of isotones shown in Fig. 6 are listed.
They include the Yb ($Z=70$) isotopes, in which the hw irrep for protons is changed from (20,0) to (10,14), as well as the $N=94$ isotones, in which the neutron irrep is changed from (36,0) to (28,10). The following observations can be made.

a) The modified irreps are quite different from the original ones, especially as far as $\mu$ is concerned.  

b) Because of a), the modified values of $\gamma$ are very different from the original ones. 

c) Despite of a), the values of the collective variable $\beta$ are changed only by a few percent. This is due to the fact that the square of the collective variable $\beta$
is proportional to the eigenvalue of the second order Casimir operator of SU(3) \cite{Iachello1987}, $C_2(\lambda,\mu)$, 
\begin{equation}
C_2(\lambda,\mu) = {2\over 3} [(\lambda+\mu)(\lambda+\mu+3)-\lambda \mu], 
\end{equation}
through the relation \cite{Castanos1988}
\begin{equation} \label{beta}
\beta^2  \propto  ((\lambda+\mu)(\lambda+\mu+3) -\lambda \mu +3),
\end{equation}      
where the proportionality constant depends of the atomic number $A$ and the dimensionless mean nuclear radius (see \cite{Bonatsos2017b} for further details). 
As one can see in Table III, the value of $\lambda+\mu$ is changed very little when passing from the original to the new irrep, thus $\beta$ remains practically unchanged. 

d) The values of $\beta$ predicted by proxy-SU(3) in a parameter-free way are in good agreement with the available experimental values \cite{Pritychenko2016}. 

As mentioned above, we approximately consider 50\% contribution from each irrep. Replacing the original values $\gamma_{hw}$ by the average of the values of $\gamma_{hw}$ and $\gamma_{nhw}$ of Table III in Fig. 6, we obtain the panels in its right column.



\begin{table*}

\caption{Nuclei with experimentally known $2^+_1$ and $2_2^+$ levels, as well as known $B(E2; 2^+_2\to 0^+_1)$ and $B(E2; 2^+_2\to 2^+_1)$ transition rates.  
Data have been taken from Ref. \cite{ensdf}.  In nuclei in which $\beta$- and $\gamma$-bands are assigned in Ref. \cite{ensdf}, the $2^+$ state of the $\gamma$-band is chosen as the $2_2^+$. Energies are given in MeV, while B(E2) transition rates are given in W.u. The ratios $R$ and $R_2$ are calculated from Eqs. (\ref{R}) and (\ref{R2}) respectively. The ratio $R_{4/2}$, a well-known \cite{Casten2000} indicator of collectivity, is also shown. See Section \ref{emp} for further discussion. 
}
\begin{tabular}{ r   r  r r c c r r }
\hline
nucleus  & $2_1^+$ & $2_2^+$ & $R$ & $2_2^+\to 0_1^+$ & $2_2^+\to 2_1^+$ & $R_2$  & $R_{4/2}$ \\ 
         & MeV     &  MeV    &     &   W.u.           &  W.u.            &        &           \\
\hline

\isotope[42][20]{Ca}$_{22}$ & 1.525 & 3.392 & 2.225 & 0.43 (9) & 6.4 (+27-15) & 14.88 & 1.805 \\
\isotope[44][20]{Ca}$_{24}$ & 1.157 & 2.657 & 2.296 & 1.70 (+20-16) & 3.6 (+12-9) & 2.12 &  1.973 \\

\isotope[44][22]{Ti}$_{22}$ & 1.083 & 2.887 & 2.665 & 0.59 (+17-12) & 2.7 (+10-8)  &  4.58 & 2.266 \\
\isotope[46][22]{Ti}$_{24}$ & 0.889 & 2.962 & 3.331 & 0.064 (16)    & 5.2 (6)      & 81.25 & 2.260 \\
\isotope[48][22]{Ti}$_{26}$ & 0.984 & 2.421 & 2.462 & 1.12 (10)     & 6.1 (+27-22) &  5.45 & 2.334 \\

\isotope[52][24]{Cr}$_{28}$ & 1.434 & 2.965 & 2.067 & 0.005 (4) & 13. (3) & 2600.00 & 1.652  \\
\isotope[54][24]{Cr}$_{30}$ & 0.835 & 2.620 & 3.138 & 0.20 (4)  &  7. (4) & 35.00   & 2.185 \\ 

\isotope[52][26]{Fe}$_{26}$ & 0.849 & 2.759 & 3.248 & 1.7 (+7-11) & 3.3 (+17-25) & 1.94   & 2.807 \\
\isotope[56][26]{Fe}$_{30}$ & 0.847 & 2.658 & 3.138 & 0.0037 (10) & 3.3 (4)      & 891.89 & 2.462 \\
\isotope[58][26]{Fe}$_{32}$ & 0.811 & 1.675 & 2.066 & 0.87 (22)   & 10. (3)      &  11.49 & 2.561 \\

\isotope[62][30]{Zn}$_{32}$ & 0.954 & 1.805 & 1.892 & 0.32 (6)       & 18. (4)         & 56.25 & 2.292 \\
\isotope[64][30]{Zn}$_{34}$ & 0.992 & 1.799 & 1.815 & 0.225 (+25-22) & 39. (4)         & 173.33 & 2.326 \\
\isotope[66][30]{Zn}$_{36}$ & 1.039 & 1.873 & 1.802 & 0.032 (12)     & 3.3 $10^2$ (13) & 10312.50 & 2.358\\
\isotope[68][30]{Zn}$_{38}$ & 1.077 & 1.883 & 1.748 & 0.85 (5)       & 28.6 (18)       & 33.65 & 2.244 \\
\isotope[70][30]{Zn}$_{40}$ & 0.885 & 1.759 & 1.988 & 0.60 (12)      & 10. (4)         &  16.67 & 2.019 \\

\isotope[64][32]{Ge}$_{32}$ & 0.902 & 1.579 & 1.751 & 0.095 (33)   & 21. (7)      & 221.05 & 2.276 \\
\isotope[66][32]{Ge}$_{34}$ & 0.957 & 1.693 & 1.769 & 0.13 (5)     & 16. (7)      & 123.08 & 2.271 \\
\isotope[68][32]{Ge}$_{36}$ & 1.016 & 1.777 & 1.750 & 0.40 (5)     & 1.0 (5)      &   2.50 & 2.233 \\
\isotope[70][32]{Ge}$_{38}$ & 1.040 & 1.708 & 1.643 & 0.50 (8)     & 64. (11)     & 128.00 & 2.071 \\
\isotope[72][32]{Ge}$_{40}$ & 0.834 & 1.464 & 1.755 &0.130 (+18-24)& 62. (+9-11)  & 476.92 & 2.072 \\
\isotope[74][32]{Ge}$_{42}$ & 0.596 & 1.204 & 2.021 & 0.71 (11)    & 43. (6)      &  60.56 & 2.457 \\
\isotope[76][32]{Ge}$_{44}$ & 0.563 & 1.108 & 1.969 & 0.74 (+8-7)  & 31.0 (+34-29)&  41.89 & 2.505 \\

\isotope[70][34]{Se}$_{36}$ & 0.945 & 1.600 & 1.694 & 0.19 (8)      &  33. (14)   & 173.68 & 2.159 \\
\isotope[74][34]{Se}$_{40}$ & 0.635 & 1.269 & 1.999 &0.80 (23)      &  48. (14)   &  60.00 & 2.148 \\
\isotope[76][34]{Se}$_{42}$ & 0.559 & 1.216 & 2.175 & 1.24 (+13-11) &44.7 (+45-38)&  36.05 & 2.380 \\
\isotope[78][34]{Se}$_{44}$ & 0.614 & 1.309 & 2.132 & 0.76 (6)      & 22.2 (18)   &  29.21 & 2.449 \\
\isotope[80][34]{Se}$_{46}$ & 0.666 & 1.449 & 2.175 & 1.33 (7)      & 18.5 (10)   &  13.91 & 2.554 \\
\isotope[82][34]{Se}$_{48}$ & 0.655 & 1.731 & 2.645 & 1.45 (21)     & 4.1 (10)    &   2.83 & 2.650 \\   

\isotope[76][36]{Kr}$_{40}$ & 0.424 & 1.222 & 2.882 &4.00 (+31-28)&1.9 (+22-14)& 0.48 & 2.440 \\
\isotope[78][36]{Kr}$_{42}$ & 0.455 & 1.148 & 2.523 & 1.7 (3)     & 5.6 (24)   & 3.29 & 2.460 \\
\isotope[80][36]{Kr}$_{44}$ & 0.617 & 1.256 & 2.037 & 0.30 (7)    & 25. (5)    & 83.33 & 2.329 \\
\isotope[82][36]{Kr}$_{46}$ & 0.777 & 1.475 & 1.899 & 0.12        &  6.9       & 57.50 & 2.344 \\
\isotope[84][36]{Kr}$_{48}$ & 0.882 & 1.898 & 2.153 & 3.0 (7)     & 13. (3)    & 4.33 & 2.376 \\
\isotope[90][36]{Kr}$_{54}$ & 0.707 & 1.362 & 1.926 & 2.8 (+6-8)  &  14. (6)   & 5.00 & 2.589 \\

\isotope[86][38]{Sr}$_{48}$ & 1.077 & 1.854 & 1.722 & 1.28 (8)      & 7.7 (11)  & 6.02 & 2.071 \\
\isotope[90][38]{Sr}$_{52}$ & 0.832 & 1.892 & 2.275 & 0.028 (+24-10)&1.7 (+15-6)&  60.71 & 1.991 \\
\isotope[92][38]{Sr}$_{54}$ & 0.815 & 1.385 & 1.699 & 0.35 (18)     &  1.9 (10) &  5.43 & 2.053 \\ 

\isotope[88][40]{Zr}$_{48}$ & 1.057 & 1.818 & 1.720 &   0.75 (7) & 6.5 (20)     & 8.67 & 2.024 \\  
\isotope[90][40]{Zr}$_{50}$ & 2.186 & 3.308 & 1.513 & 0.589 (38) & 3.5 (+15-13) & 5.94 & 1.407 \\
\isotope[92][40]{Zr}$_{52}$ & 0.935 & 1.847 & 1.977 &    3.7 (5) & 0.001 (+25-1)& 2.7 $10^{-4}$ & 1.600 \\

\isotope[92][42]{Mo}$_{50}$ & 1.510 & 3.091 & 2.048 & 2.5 (3)     & 4.3 (13)  &  1.72 & 1.512 \\
\isotope[94][42]{Mo}$_{52}$ & 0.871 & 1.864 & 2.140 &0.40 (+8-12) & 85. (5)   & 212.5 & 1.807 \\
\isotope[96][42]{Mo}$_{54}$ & 0.778 & 1.498 & 1.925 & 1.10 (11)   & 16.4 (24) &  14.91 & 2.092  \\
\isotope[98][42]{Mo}$_{56}$ & 0.787 & 1.432 & 1.819 &1.02 (+15-12)& 48. (+9-8)& 47.06 & 1.918 \\
\isotope[100][42]{Mo}$_{58}$ & 0.536 & 1.064 & 1.986& 0.62 (6)    & 52. (7)   & 83.87 & 2.121 \\

\isotope[96][44]{Ru}$_{52}$ & 0.833 & 1.931 &  2.319 & 35. (6)     & 18.4 (24)  & 0.53 & 1.823 \\  
\isotope[98][44]{Ru}$_{54}$ & 0.652 & 1.414 &  2.168 &1.04 (+17-14)& 46. (+7-6) & 44.23 & 2.142 \\
\isotope[100][44]{Ru}$_{56}$ & 0.540 & 1.362 & 2.525 & 2.0 (4)     & 31. (6)    & 15.50 & 2.273 \\
\isotope[102][44]{Ru}$_{58}$ & 0.475 & 1.103 & 2.322 & 1.14 (15)   & 32. (5)    & 28.07 & 2.329 \\
\isotope[104][44]{Ru}$_{60}$ & 0.358 & 0.893 & 2.495 & 2.8 (5)     & 55. (6)    & 19.64 & 2.482 \\

\isotope[102][46]{Pd}$_{56}$ & 0.556 & 1.534 & 2.758 & 4.2 (21)  & 15.0 (20) & 3.57 & 2.293 \\  
\isotope[104][46]{Pd}$_{58}$ & 0.556 & 1.342 & 2.414 & 1.29 (10) & 21.8 (17) & 16.90 & 2.381 \\
\isotope[106][46]{Pd}$_{60}$ & 0.512 & 1.128 & 2.204 & 1.17 (10) & 44. (4)   & 37.61 & 2.402 \\

\hline
\end{tabular}

\end{table*}

\setcounter{table}{0}

\begin{table*}

\caption{ Table I (continued) 
}
\begin{tabular}{ r r r r c c r r }
\hline
nucleus  & $2_1^+$ & $2_2^+$ & $R$ & $2_2^+\to 0_1^+$ & $2_2^+\to 2_1^+$ & $R_2$ & $R_{4/2}$ \\ 
         & MeV     &  MeV    &     &   W.u.           &  W.u.            &       &            \\
\hline

\isotope[108][46]{Pd}$_{62}$ & 0.434 & 0.931 & 2.146 & 0.83 (9)  & 72. (6)   & 86.75 & 2.416\\
\isotope[110][46]{Pd}$_{64}$ & 0.374 & 0.814 & 2.177 & 0.74 (10) & 44. (3)   & 59.46 & 2.463 \\

\isotope[106][48]{Cd}$_{58}$ & 0.633 & 1.717 & 2.713 & 2.4 (3)  & 11. (3) & 4.58   & 2.361 \\  
\isotope[108][48]{Cd}$_{60}$ & 0.633 & 1.602 & 2.531 & 1.8 (3)  & 17. (5) & 9.44   & 2.383 \\
\isotope[110][48]{Cd}$_{62}$ & 0.658 & 1.476 & 2.244 & 1.35 (20) & 30. (5) & 22.22 & 2.345 \\
\isotope[112][48]{Cd}$_{64}$ & 0.618 & 1.312 & 2.125 & 0.65 (11)& 39. (7) & 60.00  & 2.292 \\
\isotope[114][48]{Cd}$_{66}$ & 0.558 & 1.210 & 2.166 & 0.48 (6) & 22. (6) & 45.83  & 2.299 \\
\isotope[116][48]{Cd}$_{68}$ & 0.513 & 1.213 & 2.362 & 1.11 (18)& 25. (10)& 22.52  & 2.375 \\

\isotope[112][50]{Sn}$_{62}$ & 1.257 & 2.151 & 1.712 & 0.039 (12)   & 1.4 (7) & 35.90  & 1.788 \\  
\isotope[116][50]{Sn}$_{66}$ & 1.294 & 2.225 & 1.720 & 0.118 (7)    & 7.7 (8) & 65.34  & 1.848 \\
\isotope[118][50]{Sn}$_{68}$ & 1.230 & 2.120 & 1.724 & 0.075 (11)   & 6.9 (10)& 92.00  & 1.854 \\
\isotope[120][50]{Sn}$_{70}$ & 1.171 & 2.097 & 1.791 & 0.11 (4)     & 12. (4) & 109.09 & 1.873 \\
\isotope[122][50]{Sn}$_{72}$ & 1.141 & 2.154 & 1.888 & 0.015 (+5-12)& 19. (3) &1266.67 & 1.878 \\
\isotope[124][50]{Sn}$_{74}$ & 1.132 & 2.130 & 1.882 & 0.012 (+4-8) & 17.4 (4)&1450.00 & 1.857 \\

\isotope[122][52]{Te}$_{70}$ & 0.564 & 1.257 & 2.228 & 1.2 (3)     & 1.1 $10^{2}$ (3)& 91.67   & 2.094 \\  
\isotope[126][52]{Te}$_{74}$ & 0.666 & 1.420 & 2.131 & 0.140 (15)  &  44. (5)        & 314.29  & 2.043 \\
\isotope[128][52]{Te}$_{76}$ & 0.743 & 1.520 & 2.045 &0.032 (+8-16)& 28. (+7-14)     & 875.00  & 2.014 \\

\isotope[124][54]{Xe}$_{70}$ & 0.354 & 0.847 & 2.391 & 0.71 (13) & 32. (6) & 45.07 & 2.483 \\
\isotope[126][54]{Xe}$_{72}$ & 0.389 & 0.880 & 2.264 & 0.68 (12) & 47. (9) & 69.12 & 2.424 \\
\isotope[128][54]{Xe}$_{74}$ & 0.443 & 0.969 & 2.189 & 0.76 (5)  & 57. (4) & 75.00 & 2.333 \\
\isotope[132][54]{Xe}$_{78}$ & 0.668 & 1.298 & 1.944 & 0.079 (11)& 41. (4) &518.99 & 2.157 \\

\isotope[128][56]{Ba}$_{72}$ & 0.284 & 0.885 & 3.114 & 3.4 (6) & 32. (4)  & 9.41  & 2.688 \\ 
\isotope[132][56]{Ba}$_{76}$ & 0.465 & 1.032 & 2.221 & 3.9 (4) & 144. (14)&36.92  & 2.428 \\
\isotope[134][56]{Ba}$_{78}$ & 0.605 & 1.168 & 1.931 &0.42 (13)& 73. (22) &173.81 & 2.316 \\
\isotope[136][56]{Ba}$_{80}$ & 0.819 & 1.551 & 1.895 &0.89 (25)& 17. (6)  &19.10  & 2.280 \\

\isotope[136][58]{Ce}$_{78}$ & 0.552 & 1.092 & 1.978 & 0.53 (9) & 47. (8) & 88.68 & 2.380 \\

\isotope[144][60]{Nd}$_{84}$ & 0.697 & 1.561 & 2.241 & 0.210 (19) & 19.1 (23) & 90.95 & 1.887 \\

\isotope[148][62]{Sm}$_{86}$ & 0.550 & 1.454 & 2.643 & 3.3 (4) & 30. (3) & 9.09  & 2.145 \\
\isotope[152][62]{Sm}$_{90}$ & 0.122 & 1.086 & 8.916 & 2.9 (4) & 7.4 (10) & 2.55 & 3.009 \\

\isotope[154][64]{Gd}$_{90}$ & 0.123 & 0.996 &  8.095 & 5.7 (5)   & 12.3 (10)    & 2.16 & 3.015 \\ 
\isotope[156][64]{Gd}$_{92}$ & 0.089 & 1.154 & 12.972 & 4.68 (16) &  7.24 (25)   & 1.55 & 3.239 \\
\isotope[158][64]{Gd}$_{94}$ & 0.080 & 1.187 & 14.930 & 3.4 (3)   & 6.0 (7)      & 1.76 & 3.288 \\
\isotope[160][64]{Gd}$_{96}$ & 0.075 & 0.988 & 13.133 & 3.80 (16) & 1.07 (+21-16)& 0.28 & 3.302 \\

\isotope[156][66]{Dy}$_{90}$ & 0.138 & 0.891 &  6.464 & 7.2 (8)     & 9.4 (12) & 1.31 & 2.934 \\ 
\isotope[158][66]{Dy}$_{92}$ & 0.099 & 0.946 &  9.567 & 5.9 (12)    & 19. (4)  & 3.22 & 3.206 \\
\isotope[160][66]{Dy}$_{94}$ & 0.087 & 0.966 & 11.133 &4.46 (+33-99)& 8.5 (6)  & 1.91 & 3.270 \\
\isotope[162][66]{Dy}$_{96}$ & 0.081 & 0.888 & 11.011 &4.65 (+24-22)& 8.23 (41)& 1.77 & 3.294 \\
\isotope[164][66]{Dy}$_{98}$ & 0.073 & 0.762 & 10.380 & 4.3 (3)     & 7.4 (6)  & 1.72 & 3.301 \\

\isotope[162][68]{Er}$_{94}$ & 0.102 & 0.901 & 8.827 & 6.22 (+41-39)& 14.7 (9) & 2.36 & 3.230 \\
\isotope[164][68]{Er}$_{96}$ & 0.091 & 0.860 & 9.414 & 5.3 (6)      &  9. (2)  & 1.70 & 3.277 \\
\isotope[166][68]{Er}$_{98}$ & 0.081 & 0.786 & 9.753 & 5.17 (21)    & 9.6 (6)  & 1.86 & 3.289 \\

\isotope[168][70]{Yb}$_{98}$ & 0.088 & 0.984 &  11.216 & 5.0 (7)  & 9.2 (10)     & 1.84 & 3.266 \\ 
\isotope[170][70]{Yb}$_{100}$ & 0.084 & 1.146 & 13.598 & 2.7 (6)  & 4.8 (10)     & 1.78 & 3.293 \\
\isotope[176][70]{Yb}$_{106}$ & 0.082 & 1.261 & 15.351 & 1.80 (21)& 0.000111 (15)&  6.2 $10^{-5}$ & 3.310 \\
 
\isotope[174][72]{Hf}$_{102}$ & 0.091 & 1.227 & 13.483 & 4.8 (22)& 7.4 (15)& 1.54 & 3.268 \\
\isotope[178][72]{Hf}$_{106}$ & 0.093 & 1.175 & 12.606 & 3.9 (5) & 4.4 (3) & 1.13 & 3.291 \\
\isotope[180][72]{Hf}$_{108}$ & 0.093 & 1.200 & 12.855 & 3.7 (5) & 5.2 (6) & 1.41 & 3.307 \\

\isotope[182][74]{W}$_{108}$ & 0.100 & 1.221 &12.201 & 3.40 (9)     & 6.74 (17) & 1.98 & 3.291 \\
\isotope[184][74]{W}$_{110}$ & 0.111 & 0.903 & 8.122 & 4.19 (11)    & 7.94 (21) & 1.89 & 3.273 \\
\isotope[186][74]{W}$_{112}$ & 0.123 & 0.738 & 6.018 & 4.35 (+28-26)& 10.1  (7) & 2.32 & 3.234 \\

\isotope[186][76]{Os}$_{110}$ & 0.137 & 0.767 & 5.596 & 9.4 (8)      &  22.1 (17)    & 2.35 & 3.165 \\  
\isotope[188][76]{Os}$_{112}$ & 0.155 & 0.633 & 4.083 & 5.0 (6)      &  16.2 (8)     & 3.24 & 3.083 \\
\isotope[190][76]{Os}$_{114}$ & 0.187 & 0.558 & 2.988 & 6.0 (6)      & 32.6 (34)     & 5.43 & 2.934 \\
\isotope[192][76]{Os}$_{116}$ & 0.206 & 0.489 & 2.376 & 5.62 (+21-12)&  46.0 (+26-12)& 8.19 & 2.820 \\

\isotope[192][78]{Pt}$_{114}$ & 0.317 & 0.612 & 1.935 & 0.55 (4)       & 109. (7)     & 198.18 & 2.479 \\  
\isotope[194][78]{Pt}$_{116}$ & 0.328 & 0.622 & 1.894 & 0.286 (+44-35) & 89. (+12-10) & 311.19 & 2.470 \\
\isotope[198][78]{Pt}$_{120}$ & 0.407 & 0.775 & 1.895 & 0.038 (12)     & 37. (7)      & 973.68 & 2.419 \\

\isotope[198][80]{Hg}$_{118}$ & 0.412 & 1.088 & 2.641 & 0.0216 (4) & 0.63 (8) & 29.17 & 2.546 \\ 
\isotope[200][80]{Hg}$_{120}$ & 0.368 & 1.254 & 3.408 & 0.23 (5)   & 2.2 (5)  &  9.57 & 2.574 \\
\isotope[202][80]{Hg}$_{122}$ & 0.440 & 0.960 & 2.184 & 0.087 (21) & 5.6 (15) & 64.37 & 2.548 \\

\hline
\end{tabular}

\end{table*}


\begin{table*}

\caption{ Highest weight (hw) irreducible representations (irreps) and next higher weight (nhw) irreps for the algebras U(6), U(10), U(15), U(21), and U(28) corresponding to the $sd$, $pf$, $sdg$, $pfh$, and $sdgi$ shells of the isotropic 3-dimensional harmonic oscillator, as well as to the shell model 8-20, 28-50, 50-82, 82-126, and 126-184 shells within the proxy-SU(3) scheme \cite{Martinou2020}, calculated using the code UNTOSU3 \cite{Draayer1989}. See Appendix A for further discussion. 
}
\begin{tabular}{ r r r r r r r r r r r  }
\hline
   & U(6) & U(6) & U(10) & U(10)& U(15) & U(15) & U(21) & U(21) & U(28) & U(28) \\ 
   & $sd$ & $sd$ & $pf$  & $pf$ & $sdg$ & $sdg$ & $pfh$ & $pfh$ & $sdgi$ & $sdgi$ \\
   & 8-20 & 8-20 & 28-50 & 28-50 & 50-82 & 50-82 & 82-126 & 82-126 & 126-184 & 126-184 \\
M  & hw & nhw & hw & nhw & hw &  nhw & hw & nhw & hw & nhw \\  
\hline
 0 &  0,0  &       & 0,0  &       &   0,0 &       &   0,0 &       &  0,0  &      \\
 2 &  4,0  &  0,2  & 6,0  &  2,2  &   8,0 &   4,2 &  10,0 &   6,2 &  12,0 &  8,2 \\
 4 &  4,2  &  0,4  & 8,2  &  4,4  &  12,2 &   8,4 &  16,2 &  12,4 &  20,2 & 16,4 \\
 6 &  6,0  &  0,6  & 12,0 &  6,6  &  18,0 &  12,6 &  24,0 &  18,6 &  30,0 & 24,6 \\
 8 &  2,4  &  4,0  & 10,4 & 12,0  &  18,4 &  20,0 &  26,4 &  28,0 &  34,4 & 36,0 \\
10 &  0,4  &  2,0  & 10,4 & 12,0  &  20,4 &  22,0 &  30,4 &  32,0 &  40,4 & 42,0 \\
12 &  0,0  &       & 12,0 &  4,10 &  24,0 & 16,10 &  36,0 & 28,10 &  48,0 & 40,10 \\
14 &       &       & 6,6  &  8,2  &  20,6 &  22,2 &  34,6 &  36,2 &  48,6 & 50,2 \\
16 &       &       & 2,8  &  4,4  &  18,8 &  20,4 &  34,8 &  36,4 &  50,8 & 52,4 \\   
18 &       &       & 0,6  &  2,2  &  18,6 &  20,2 &  36,6 &  38,2 &  54,6 & 56,2 \\
20 &       &       & 0,0  &       &  20,0 & 10,14 &  40,0 & 30,14 &  60,0 & 50,14 \\
22 &      &        &      &       &  12,8 &  14,4 &  34,8 &  36,4 &  56,8 & 58,4 \\
24 &      &        &      &       &  6,12 &   8,8 & 30,12 &  32,8 & 54,12 & 56,8 \\
26 &      &        &      &       &  2,12 &   4,8 & 28,12 &  30,8 & 54,12 & 56,8 \\
28 &      &        &      &       &   0,8 &   2,4 &  28,8 &  30,4 &  56,8 & 58,4 \\
30 &      &        &      &       &   0,0 &       &  30,0 & 18,18 &  60,0 & 48,18 \\
32 &      &        &      &       &       &       & 20,10 &  22,6 & 52,10 & 54,6 \\
34 &      &        &      &       &       &       & 12,16 & 14,12 & 46,16 & 48,12 \\
36 &      &        &      &       &       &       &  6,18 &  8,14 & 42,18 & 44,14 \\
38 &      &        &      &       &       &       &  2,16 &  4,12 & 40,16 & 42,12 \\
40 &      &        &      &       &       &       &  0,10 &   2,6 & 40,10 & 42,6  \\
42 &      &        &      &       &       &       &   0,0 &       & 42,0  & 28,22\\
44 &      &        &      &       &       &       &       &       & 30,12 & 32,8   \\
46 &      &        &      &       &       &       &       &       & 20,20 & 22,16 \\
48 &      &        &      &       &       &       &       &       & 12,24 & 14,20 \\
50 &      &        &      &       &       &       &       &       &  6,24 &  8,20\\
52 &      &        &      &       &       &       &       &       &  2,20 &  4,16 \\
54 &      &        &      &       &       &       &       &       &  0,12 &  2,8 \\
56 &      &        &      &       &       &       &       &       &  0,0  &      \\   

\hline
\end{tabular}

\end{table*}


\begin{table}

\caption{Nuclei of Fig. 6 having $Z=62$, 70 and/or $N=94$, for which the next higher weight (nhw) irrep of Table II has to be taken into account. In column hw the proxy-SU(3) irreps obtained from the hw valence protons and neutrons irreps are shown, while in column nhw the proxy-SU(3) irreps resulting after the replacement of hw irreps having $\mu=0$ by their nhw irreps are given. The values of the collective variables $\gamma$ and $\beta$ obtained with these irreps from Eqs. (\ref{mu}) and (\ref{beta}) are labeled by the subscripts hw and nhw respectively. Experimental values of $\beta$ \cite{Pritychenko2016}, labeled by exp, are also given for comparison.   See Appendix B for further discussion. 
}
\begin{tabular}{ r r r r r r r c  }
\hline
nucleus  & hw & nhw & $\gamma_{hw}$ & $\gamma_{nhw}$ & $\beta_{hw}$ & $\beta_{nhw}$ & $\beta_{exp}$ \\ 
         &    &         &   deg         &         deg        &              &        &           \\           
\hline

\isotope[154][60]{Nd}$_{94}$ & 56,4 & 48,14 & 4.16 & 12.95 & 0.307 & 0.298 & 0.258 (+42-58) \\
\isotope[156][62]{Sm}$_{94}$ & 60,0 & 44,20 & 0.81 & 18.14 & 0.315 & 0.299 & $<0.434$ \\
\isotope[158][64]{Gd}$_{94}$ & 56,6 & 48,16 & 5.72 & 14.36 & 0.310 & 0.303 & 0.3510 (38) \\
\isotope[160][66]{Dy}$_{94}$ & 54,8 & 46,18 & 7.46 & 16.24 & 0.305 & 0.299 & 0.3360 (13) \\
\isotope[162][68]{Er}$_{94}$ & 54,6 & 46,16 & 5.92 & 14.86 & 0.297 & 0.290 & 0.3232 (81) \\
\isotope[164][70]{Yb}$_{94}$ & 56,0 & 38,24 & 0.86 & 22.80 & 0.290 & 0.281 & 0.2886 (47) \\
\isotope[166][72]{Hf}$_{94}$ & 48,8 & 40,18 & 8.29 & 18.05 & 0.271 & 0.266 & 0.2488 (+61-54) \\

\isotope[166][70]{Yb}$_{96}$ & 54,6 & 44,20 & 5.92 & 18.14 & 0.294 & 0.293 & 0.3137 (60) \\
  
\isotope[168][70]{Yb}$_{98}$ & 54,8 & 44,22 & 7.46 & 19.42 & 0.300 & 0.299 & 0.324 (3) \\

\hline
\end{tabular}

\end{table}


\begin{figure*} [htb]

    \includegraphics[width=185mm]{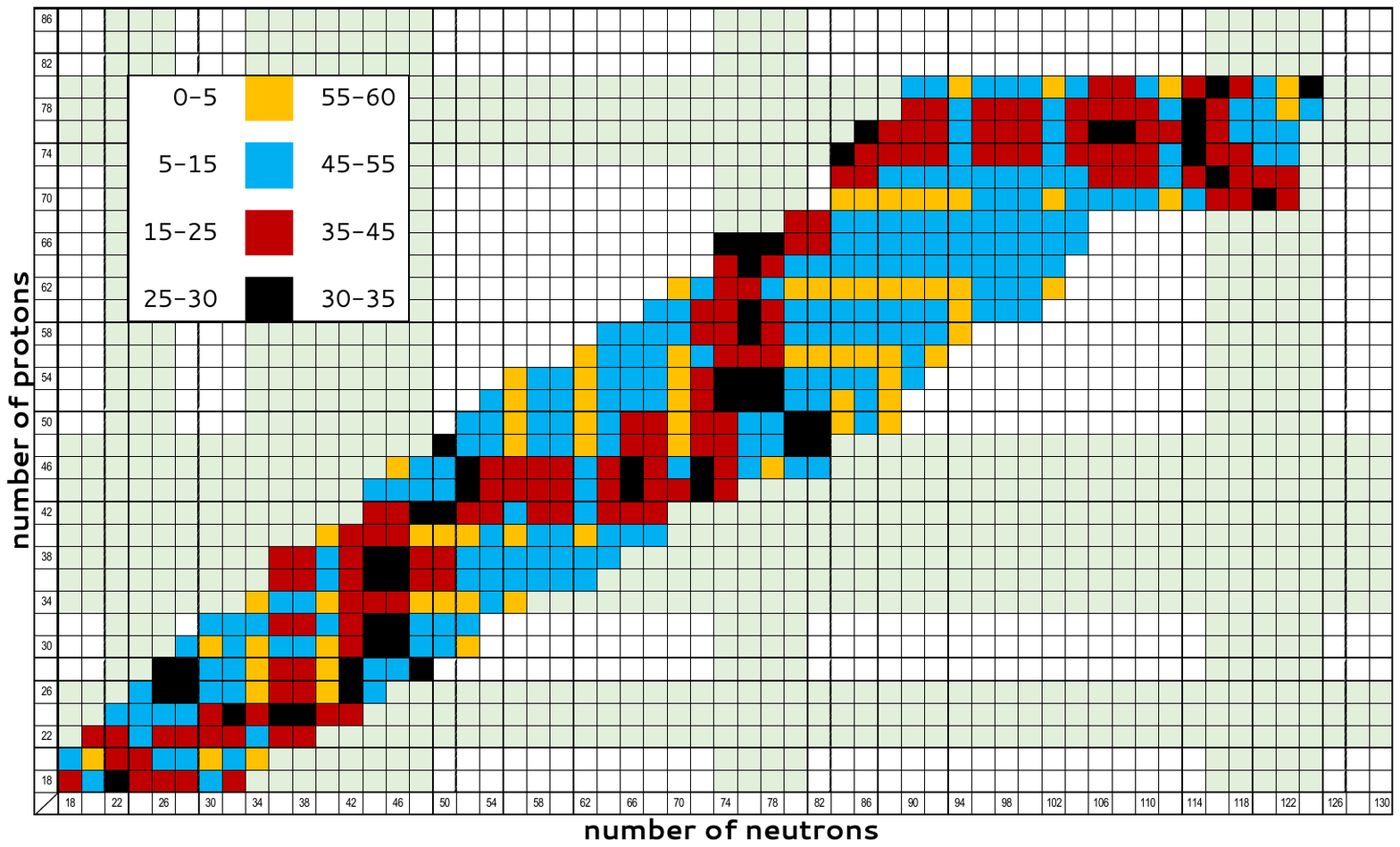}

    \caption{Proxy-SU(3) predictions, taken from Eq. (\ref{mu}),  for the deformation variable $\gamma$ (in degrees) for all experimentally known \cite{ensdf} nuclei with $18\leq Z \leq 80$ and $N\leq 18 \leq 124$.   See Sec. \ref{proxy} for further discussion.} 
    
\end{figure*}


\begin{figure*} [htb]

    \includegraphics[width=185mm]{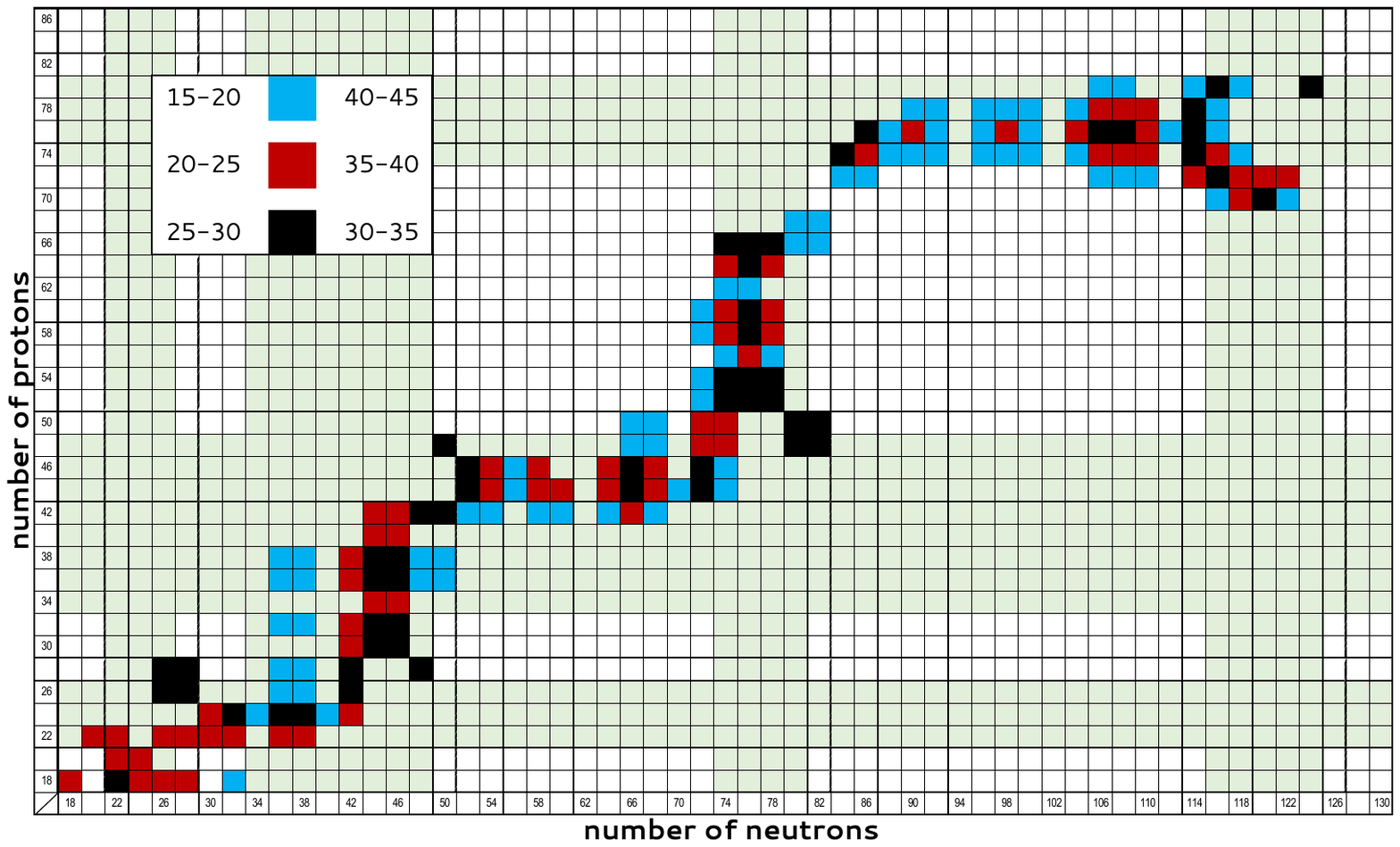}

    \caption{Same as Fig. 1, but only for nuclei having $15^{\rm o} \leq \gamma \leq 45^{\rm o} $ included. The horizontal and vertical stripes covering the nucleon numbers 22-26, 34-48, 74-80, 116-124, 172-182, within which substantial triaxiality is expected to occur, are also shown.  See Sec. \ref{proxy} for further discussion.} 
    
\end{figure*}



\begin{figure} [htb]

\includegraphics[width=75mm]{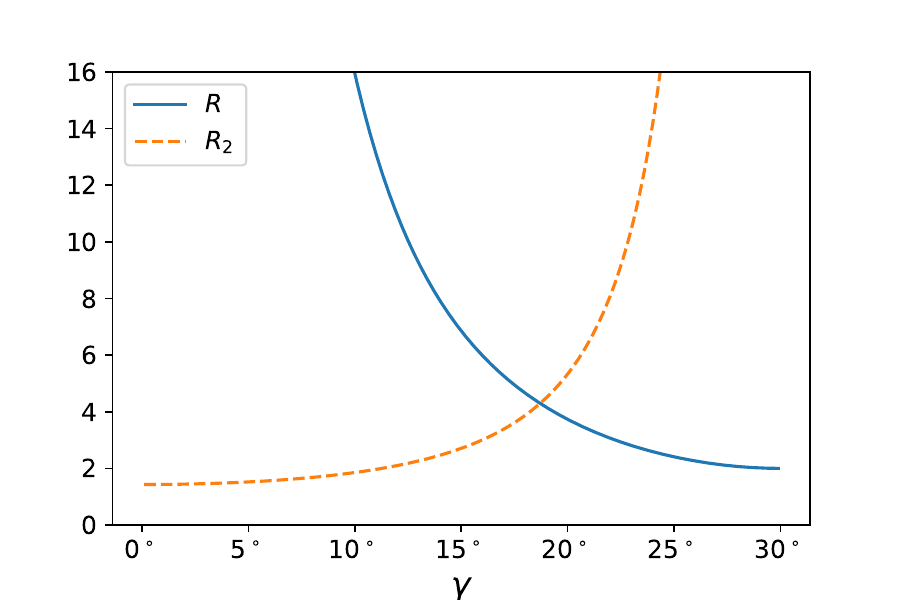}

    \caption{ The theoretical energy ratio $R$ (Eq. (\ref{R})), shown by a blue solid line, and the theoretical branching ratio $R_2$ (Eq. (\ref{R2})), shown by an orange  dashed line, are plotted as a function of $\gamma$. These ratios are used \cite{Casten2000} for singling out from the data nuclei with substantial ($\gamma\geq 15^{\rm o}$) triaxiality.  See Sec. \ref{emp} for further discussion.} 
    
\end{figure}


\begin{figure*} [htb]

    \includegraphics[width=185mm]{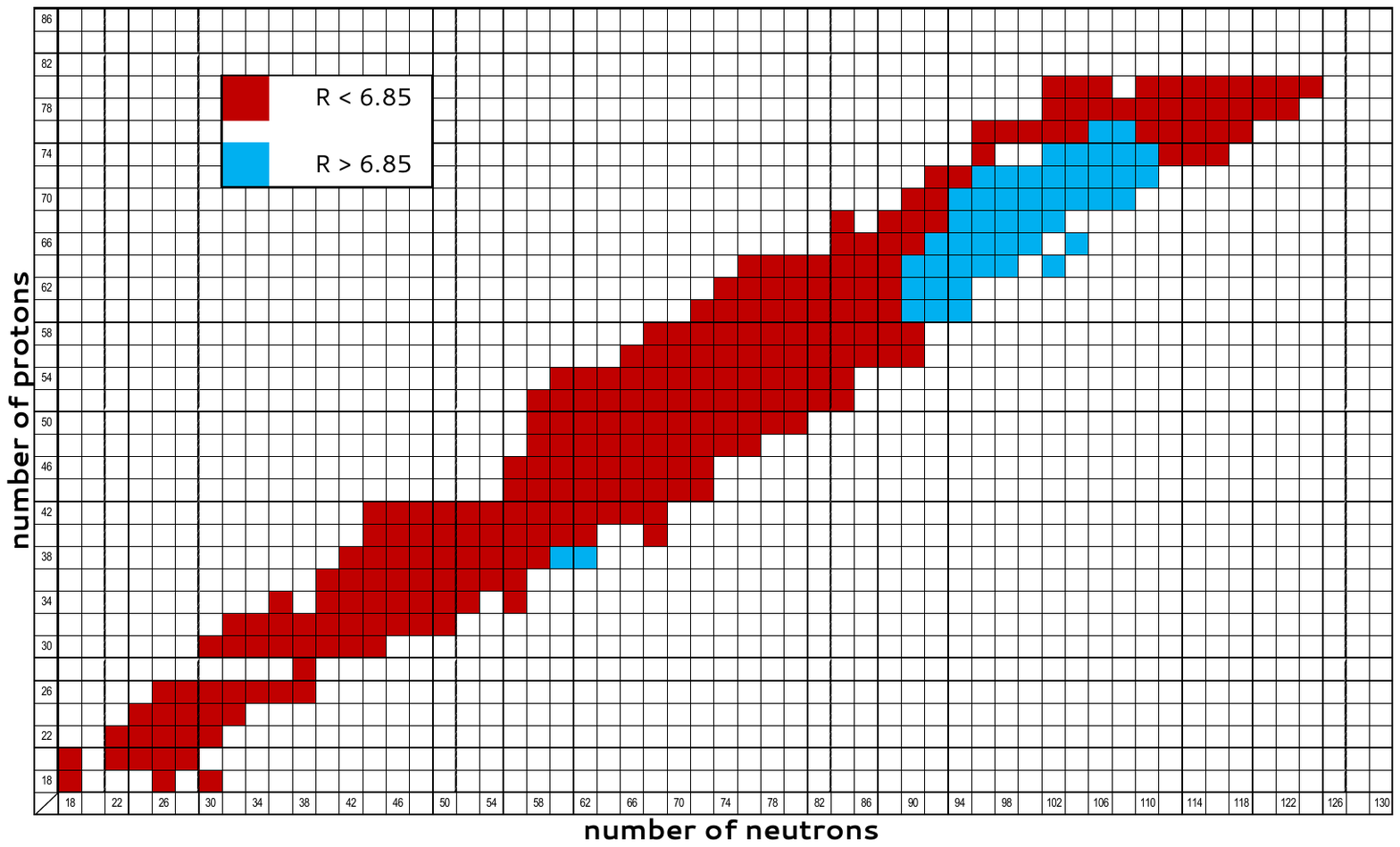}

    \caption{Nuclei with experimentally known $2_1^+$ and $2_2^+$ levels, subdivided into these with $R<6.85$ (expected to have $\gamma > 15^{\rm o}$) and those with $R>6.85$ (expected to have $\gamma < 15^{\rm o}$). Data have been taken from Ref. \cite{ensdf}. In nuclei in which $\beta$- and $\gamma$-bands are assigned in Ref. \cite{ensdf}, the $2^+$ state of the $\gamma$-band is chosen as the $2_2^+$.    See Sec. \ref{emp} for further discussion.} 
    
\end{figure*}


\begin{figure*} [htb]

    \includegraphics[width=185mm]{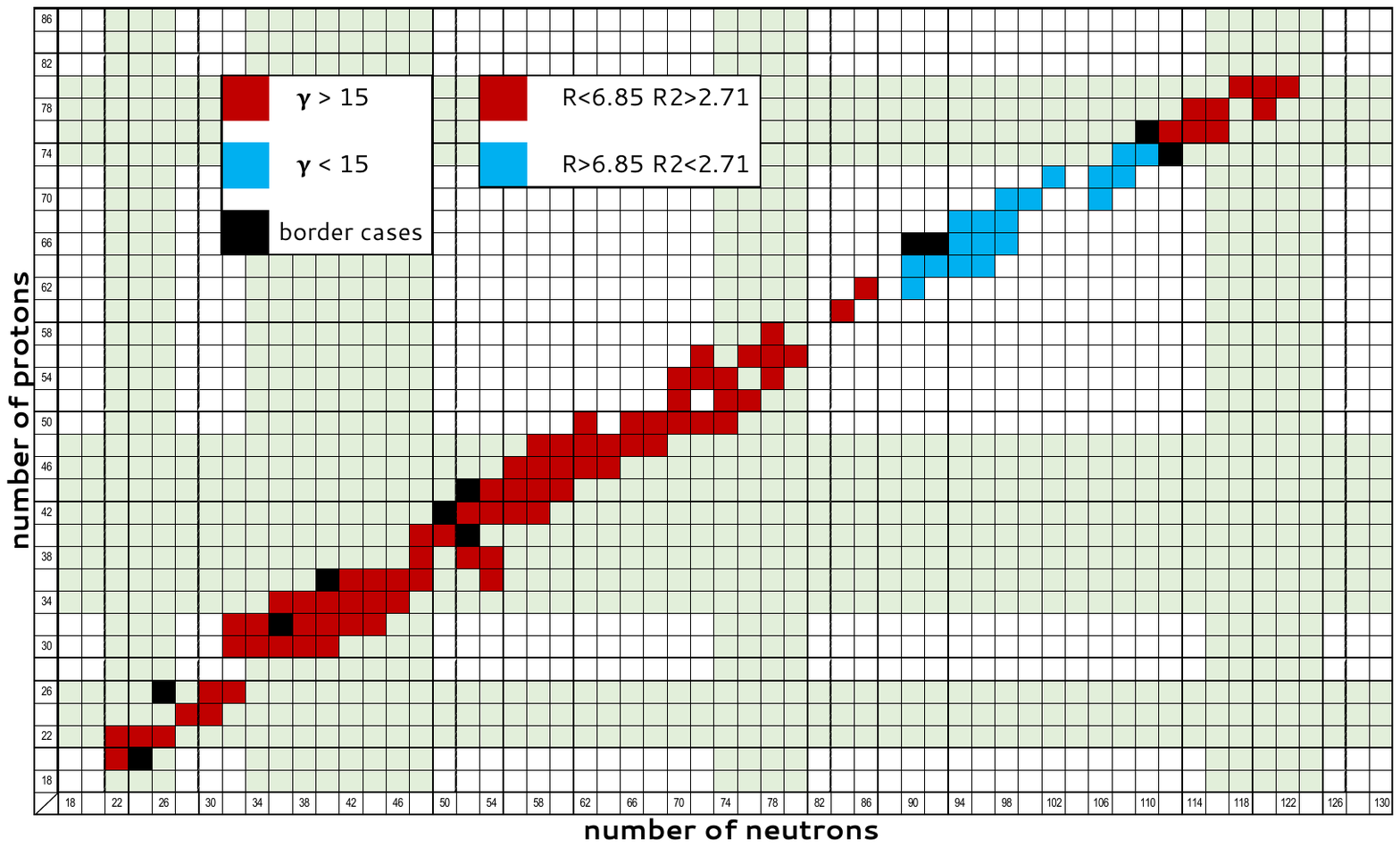}

    \caption{Nuclei with experimentally known $2_1^+$ and $2_2^+$ levels, as well as known $B(E2; 2_2^+\to 0_1^+)$ and $B(E2; 2_2^+\to 2_1^+)$ transition rates,  subdivided into these with $R<6.85$ and $R_2>2.71$ (expected to have $\gamma > 15^{\rm o}$) and those with $R>6.85$ and $R_2<2.71$ (expected to have $\gamma < 15^{\rm o}$).  
 Data have been taken from Ref. \cite{ensdf} and collected in Table I. In nuclei in which $\beta$- and $\gamma$-bands are assigned in Ref. \cite{ensdf}, the $2^+$ state of the $\gamma$-band is chosen as the $2_2^+$. Ten border-line nuclei from Table I with $R<6.85$ and $R_2<2.71$, as well as one nucleus (\isotope[158][66]{Dy}) with $R>6.85$ and $R_2>2.71$ are also shown. The nuclei expected to have $\gamma > 15^{\rm o}$ lie within the horizontal and vertical stripes predicted by the proxy-SU(3) symmetry, covering the nucleon numbers  22-26, 34-48, 74-80, 116-124, 172-182, also shown in Fig. 2.  See Sec. \ref{emp} for further discussion.} 
    
\end{figure*}


\begin{figure*} [htb]

    {\includegraphics[width=75mm]{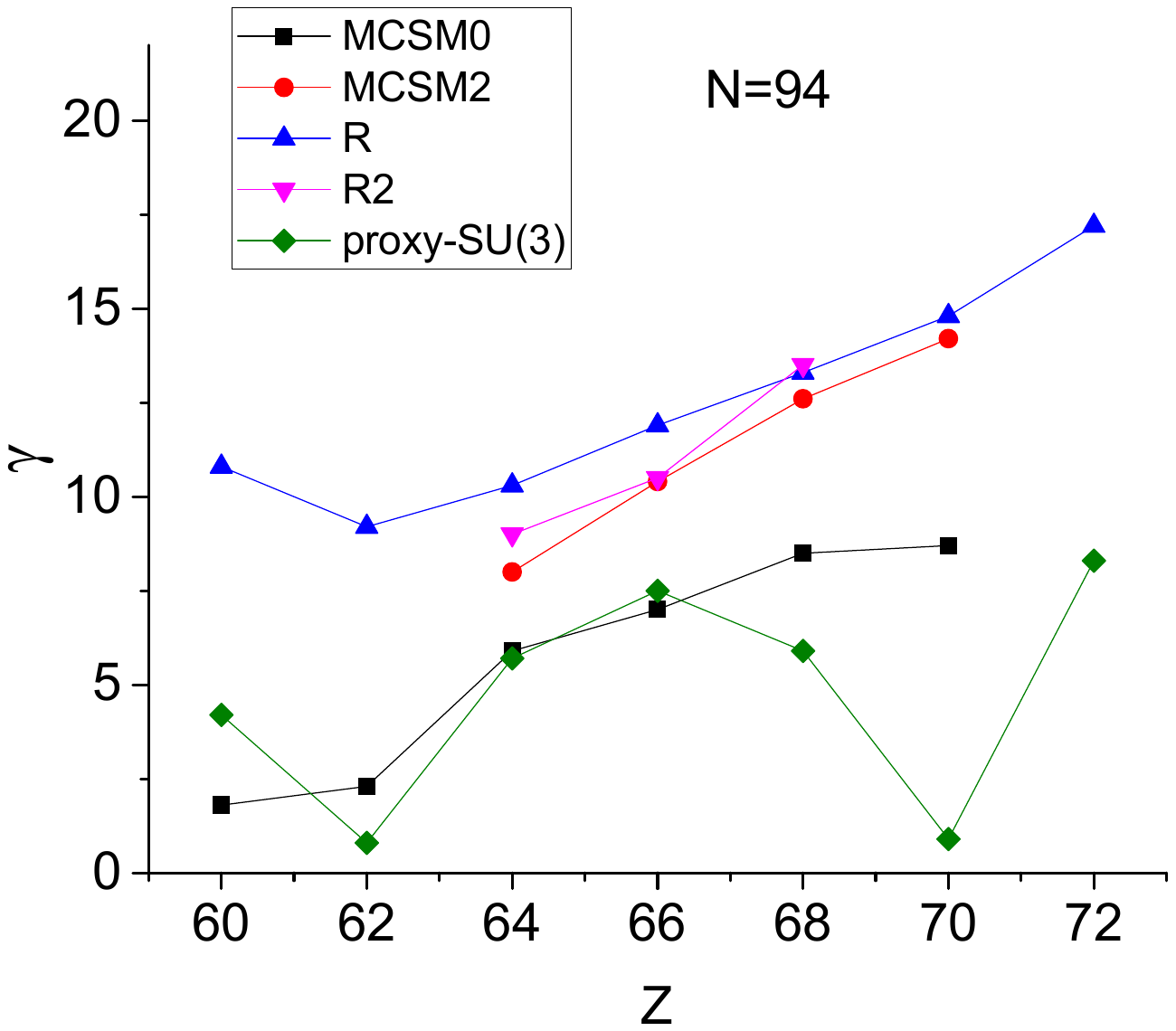} \hspace{5mm}   \includegraphics[width=75mm]{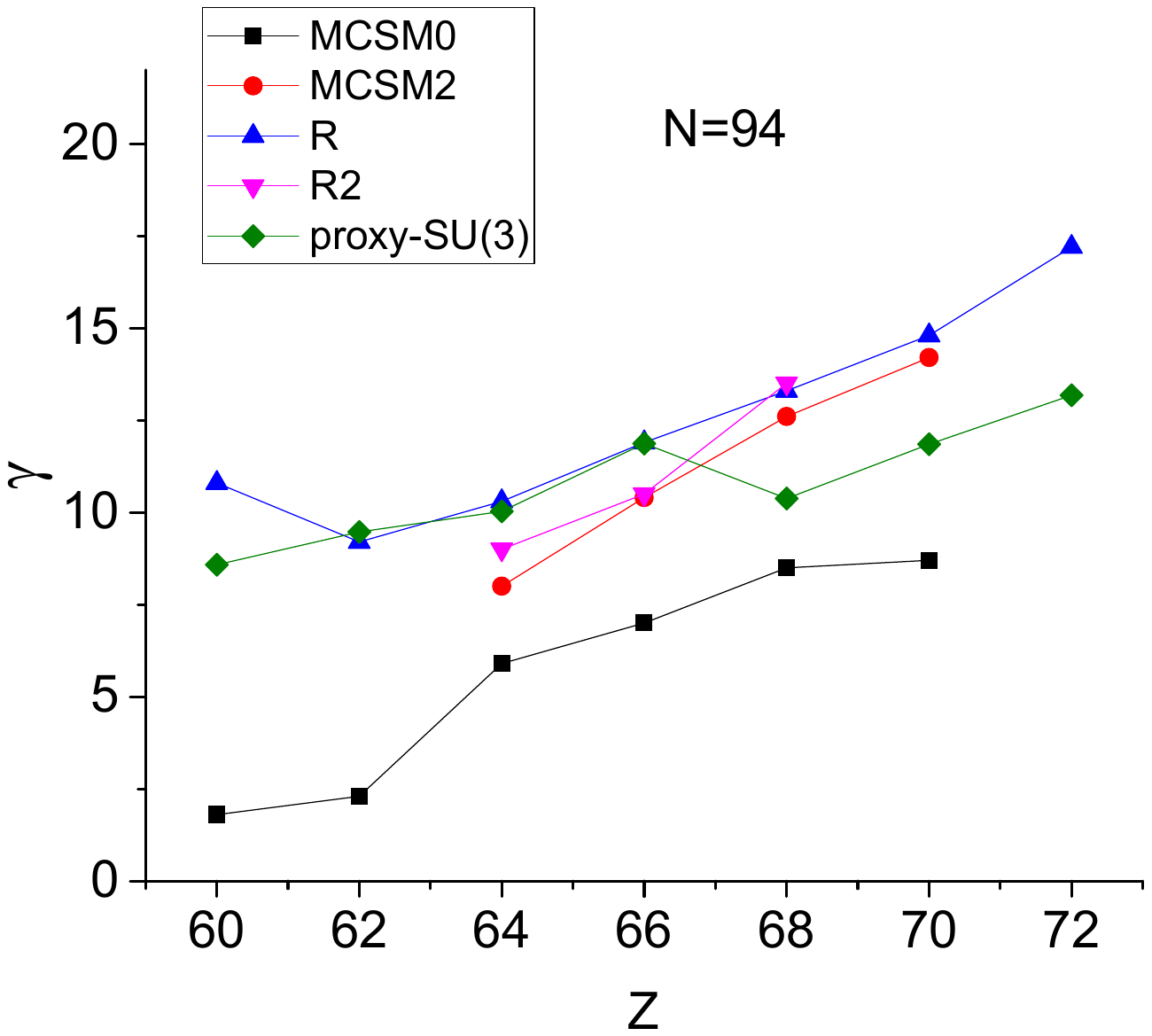} }
    {\includegraphics[width=75mm]{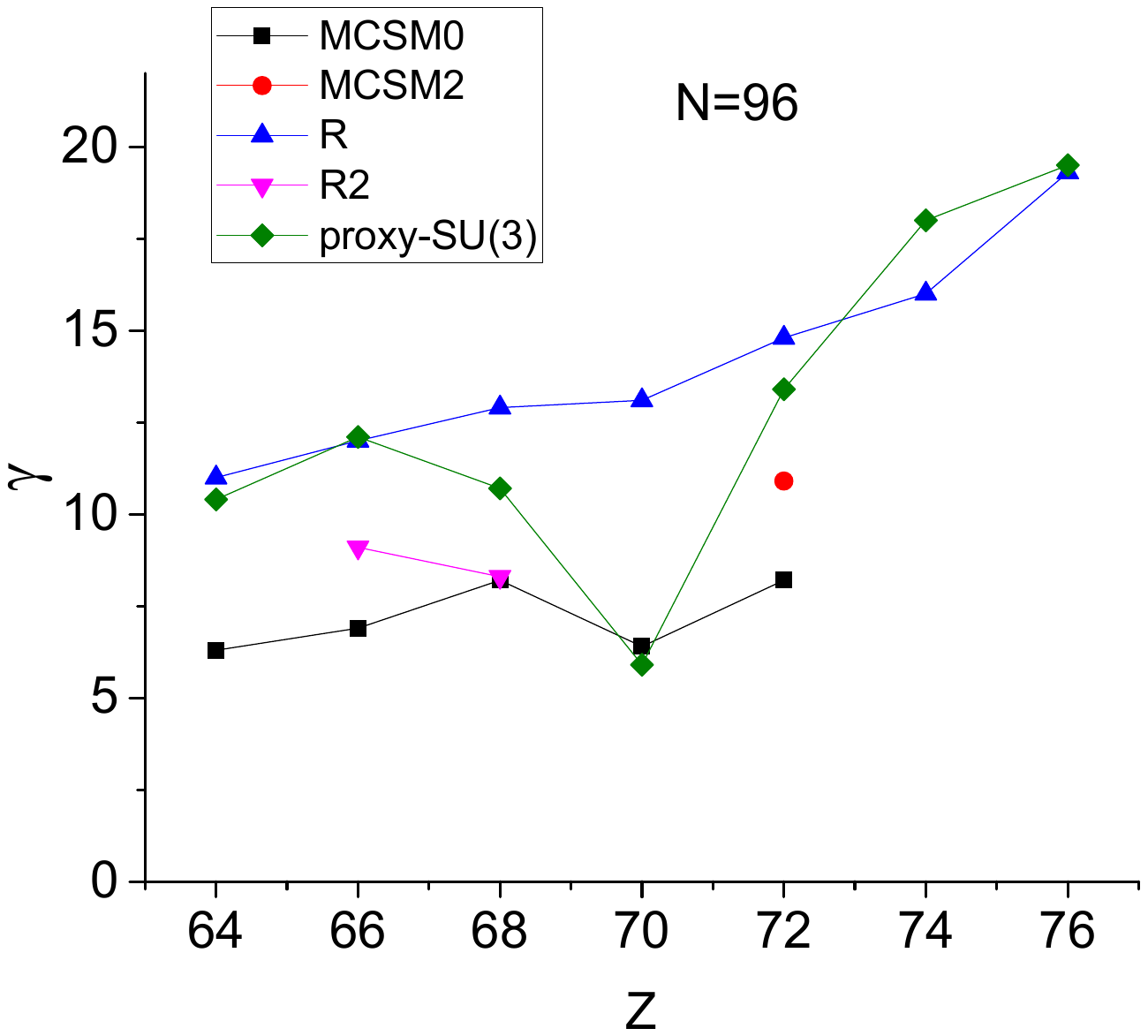} \hspace{5mm}    \includegraphics[width=75mm]{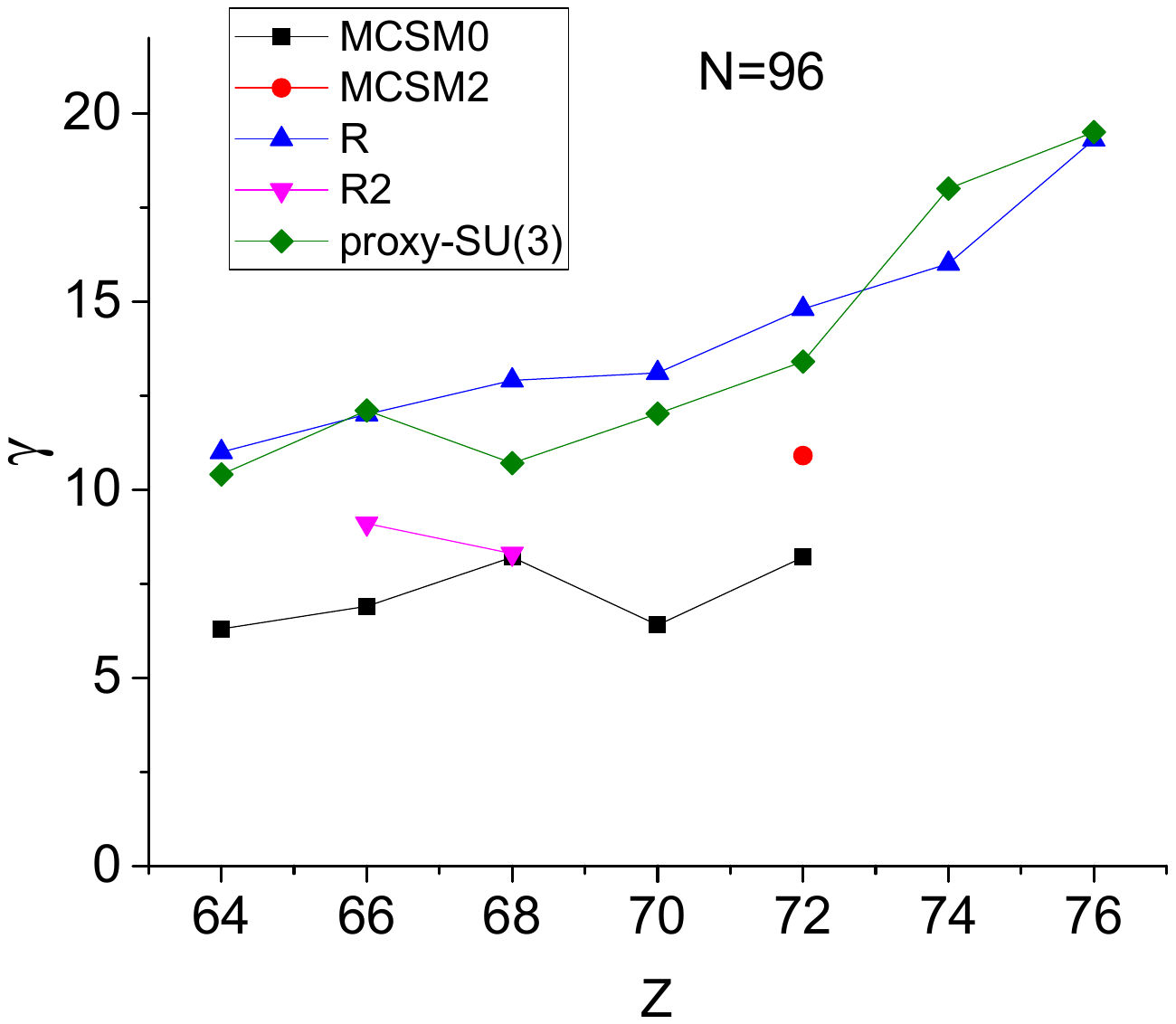} }
    {\includegraphics[width=75mm]{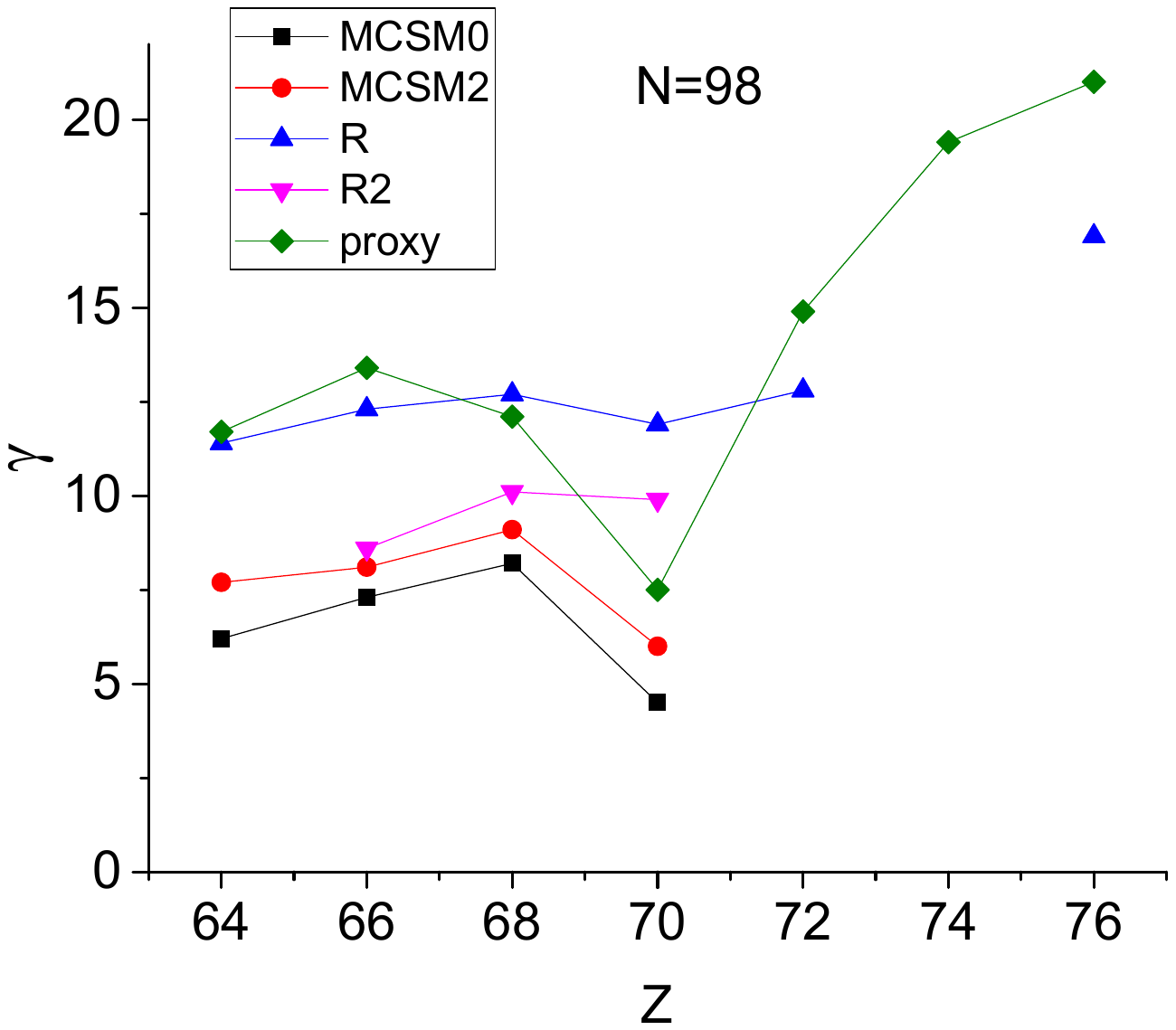}  \hspace{5mm}   \includegraphics[width=75mm]{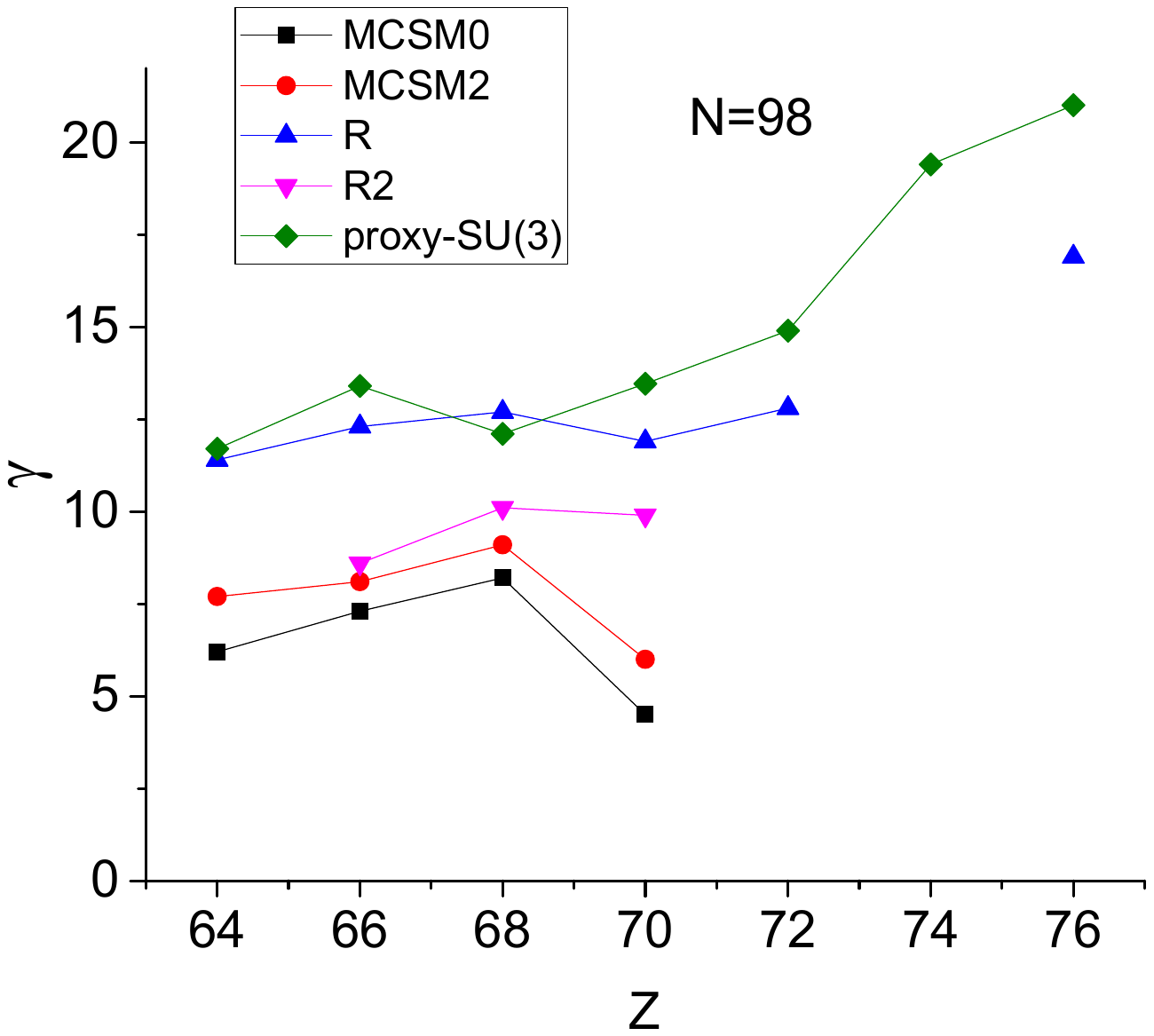} }
     
    \caption{Left column: Parameter-free proxy-SU(3) predictions (labeled as proxy-SU(3)) for $\gamma$ (calculated with Eq. (\ref{mu})) for some series of isotones are compared to the empirical values extracted from the ratios $R$ (Eq. (\ref{R})) and $R_2$ (Eq. (\ref{R2})) (labeled as $R$ and $R_2$ respectively), in the way described in Sec. \ref{emp} and depicted in Fig. 3, as well as to the predictions coming from state-of-the-art configuration interaction calculations in the Monte Carlo shell model (MCSM) framework \cite{Otsuka2024}, in which predictions for the ground state ($K=0$) band are labeled by MCSM0, while these for the lowest $K=2$ band are labeled as MCSM2. While the overall agreement is in general good, the shortcomings of proxy-SU(3) for $Z=70$ and $N=94$ become evident, calling for taking into account the next higher weight (nhw) irreps neighboring the highest weight irrep, if the latter happens to be fully symmetric (i.e., has $\mu=0$).  Right column: In this column the proxy-SU(3) predictions have been replaced by the average value of the hw and nhw values obtained in Table III, as discussed in Appendix B. See Sec. \ref{MCSM} for further discussion.} 
    
\end{figure*}


\begin{thebibliography}{999}

\bibitem{Bohr1952}
Bohr A
1952 \textit{Dan. Mat. Fys. Medd.} \textbf{26} no. 14 
The coupling of nuclear surface oscillations to the motion of individual nucleons

\bibitem{Bohr1998b}
Bohr A and Mottelson B R
1988  \textit{Nuclear Structure Vol. II: Nuclear Deformations} (Singapore: World Scientific) 

\bibitem{Davydov1958}
Davydov A S  and  Filippov G F
1958 \textit{ Nucl. Phys.} \textbf{ 8} 237
Rotational states in even atomic nuclei

\bibitem{Davydov1959}
Davydov A S and Rostovsky V S 
1959 \textit{ Nucl. Phys.} \textbf{  12} 58 
Relative transition probabilities between rotational levels of non-axial nuclei

\bibitem{MeyerterVehn1975}
Meyer-Ter-Vehn J 
1975 \textit{ Nucl. Phys. A} \textbf{  249} 111 
Collective model description of transitional odd-A nuclei: (I). The triaxial-rotor-plus-particle model

\bibitem{Arima1975}
Arima A and  Iachello F 
1975 \textit{Phys. Rev. Lett.} \textbf{35} 1069 
Collective Nuclear States as Representations of a SU(6) Group

\bibitem{Iachello1987}
Iachello F and Arima A 1987 \textit{The Interacting Boson Model} (Cambridge: Cambridge U. Press)  

\bibitem{Dieperink1982}
Dieperink A E L and  Bijker R
1982 \textit{ Phys. Lett. B} \textbf{  116} 77
On triaxial features in the neutron-proton IBA

\bibitem{Walet1987}
Walet N R and Brussaard P J 
1987 \textit{ Nucl. Phys. A} \textbf{  474}  61
A study of the SU(3)$^*$ limit of IBM-2

\bibitem{Giraud1969}
Giraud B and Sauer P U 
1969 \textit{ Phys. Lett. B} \textbf{  30} 218
Restoration of rotational symmetry for triaxial Hartree-Fock solutions

\bibitem{Girod1978}
Girod M and Grammaticos B
1978 \textit{ Phys. Rev. Lett.} \textbf{  40} 361 
Quest for Triaxial Nuclei: Some Hartree-Bogoliubov Predictions

\bibitem{Girod1983}
Girod M  and Grammaticos B
1983 \textit{ Phys. Rev. C} \textbf{  27} 2317
Triaxial Hartree-Fock-Bogolyubov calculations with D1 effective interaction

\bibitem{Hayashi1984}
Hayashi A, Hara K and Ring P 
1984 \textit{ Phys. Rev. Lett.} \textbf{  53} 337 
Existence of Triaxial Shapes in Transitional Nuclei

\bibitem{Yao2010}
Yao J M, Meng J, Ring P and Vretenar D
2010 \textit{ Phys. Rev. C} \textbf{  81} 044311 
Configuration mixing of angular-momentum-projected triaxial relativistic mean-field wave functions

\bibitem{Mayer1949}
Mayer M G
1949 \textit{Phys. Rev.} \textbf{75} 1969
On Closed Shells in Nuclei. II.

\bibitem{Haxel1949}
Haxel O,  Jensen J H D and Suess H E 
1949 \textit{Phys. Rev.} \textbf{75} 1766 
On the "Magic Numbers" in Nuclear Structure.

\bibitem{Mayer1955}
Mayer M G and  Jensen J H D 1955 \textit{Elementary Theory of Nuclear Shell Structure} (New York: Wiley) 1955. 

\bibitem{Nilsson1955}
Nilsson S G
1955 \textit{Dan. Mat. Fys. Medd.} \textbf{29} no. 16 
Binding states of individual nucleons in strongly deformed nuclei. 

\bibitem{Nilsson1995}
Nilsson S G and Ragnarsson I 1995 \textit{Shapes and Shells in Nuclear Structure} (Cambridge:  Cambridge U. Press) 

\bibitem{Sheikh1999}
Sheikh J A and Hara K 
1999 \textit{ Phys. Rev. Lett.} \textbf{  82} 3968 
Triaxial Projected Shell Model Approach

\bibitem{Rouoof2024}
Rouoof S P, Nazir N, Jehangir S, Bhat G H, Sheikh J A, Rather N and Frauendorf S
2024 \textit{ Eur. Phys. J. A} \textbf{  60} 40
Fingerprints of the triaxial deformation from energies and B(E2) transition probabilities of $\gamma$-bands in transitional and deformed nuclei

\bibitem{Honma1995}
Honma M, Mizusaki T and Otsuka T
1995 \textit{ Phys. Rev. Lett.} \textbf{  75} 1284 
Diagonalization of Hamiltonians for Many-Body Systems by Auxiliary Field Quantum Monte Carlo Technique

\bibitem{Tsunoda2021}
Tsunoda Y and  Otsuka T
2021 \textit{ Phys. Rev. C} \textbf{  103} L021303 
Triaxial rigidity of  $^{166}$Er  and its Bohr-model realization

\bibitem{Otsuka2024}
Otsuka T, Tsunoda Y, Utsuno Y, Shimizu N,  Abe T and Ueno H
2024 arXiv: 2303.11299v5 [nucl-th] 
Prevailing Triaxial Shapes in Heavy Nuclei Driven by Nuclear Tensor Force

\bibitem{Elliott1958a}
Elliott J P 
1958 \textit{Proc. Roy. Soc. A Ser. A} \textbf{245} 128 
Collective motion in the nuclear shell model. I. Classification schemes for states of mixed configurations.

\bibitem{Elliott1958b}
Elliott J P 
1958 \textit{Proc. Roy. Soc. A Ser. A} \textbf{245}  562
Collective motion in the nuclear shell model II. The introduction of intrinsic wave-functions.

\bibitem{Elliott1963}
Elliott J P and  Harvey M
1963 \textit{Proc. Roy. Soc. A Ser. A} \textbf{272}  557
Collective motion in the nuclear shell model III. The calculation of spectra.

\bibitem{Wybourne1974}
Wybourne B G 1974 \textit{Classical Groups for Physicists}  (New York: Wiley) 

\bibitem{RatnaRaju1973}
Ratna Raju  R D, Draayer J P and Hecht K T 
1973 \textit{Nucl. Phys. A} \textbf{202} 433  
Search for a coupling scheme in heavy deformed nuclei: The pseudo SU(3) model.

\bibitem{Draayer1983}
Draayer J P and Weeks K J
1983 \textit{Phys. Rev. Lett.} \textbf{51} 1422 
Shell-Model Description of the Low-Energy Structure of Strongly Deformed Nuclei.

\bibitem{Draayer1984}
Draayer J P and  Weeks K J
1984 \textit{Ann. Phys. (NY)} \textbf{156} 41 
Towards a shell model description of the low-energy structure of deformed nuclei I. Even-even systems.

\bibitem{Zuker1995}
Zuker A P, Retamosa J,  Poves A and Caurier E 
1995 \textit{Phys. Rev. C} \textbf{52}  R1741(R) 
Spherical shell model description of rotational motion.

\bibitem{Zuker2015}
Zuker A P, Poves A,  Nowacki F, Lenzi S M 
2015 \textit{Phys. Rev. C} \textbf{92} 024320 
Nilsson-SU3 self-consistency in heavy N=Z nuclei

\bibitem{Bonatsos2017a}
Bonatsos D,  Assimakis I E,  Minkov N,  Martinou A,  Cakirli R B,  Casten R F and  Blaum K 
2017 \textit{Phys. Rev. C} \textbf{95} 064325
Proxy-SU(3) symmetry in heavy deformed nuclei.

\bibitem{Bonatsos2017b}
Bonatsos D, Assimakis I E,  Minkov N, Martinou A, Sarantopoulou S, Cakirli R B, Casten R F and Blaum K 
2017 \textit{ Phys. Rev. C} \textbf{95} 064326 
Analytic predictions for nuclear shapes, prolate dominance, and the prolate-oblate shape transition in the proxy-SU(3) model

\bibitem{Bonatsos2023b}
Bonatsos D, Martinou A, Peroulis S K,  Mertzimekis T J and Minkov N
2023 \textit{Symmetry} \textbf{15} 169 
The Proxy-SU(3) Symmetry in Atomic Nuclei.

\bibitem{Castanos1988}
Casta\~{n}os O,  Draayer J P and Leschber Y
1988 \textit{Z. Phys. A} \textbf{329} 33
Shape variables and the shell model.

\bibitem{Bhat2014}
Bhat G H, Dar W A, Sheikh J A and Sun Y
2014 \textit{ Phys. Rev. C} \textbf{  89} 014328 
Nature of $\gamma$ deformation in Ge and Se nuclei and the triaxial projected shell model description

\bibitem{Zhang2015b}
Zhang C L,  Bhat G H, Nazarewicz W, Sheikh J A, and  Shi Y
2015 \textit{Phys. Rev. C} \textbf{  92} 034307 
Theoretical study of triaxial shapes of neutron-rich Mo and Ru nuclei

\bibitem{Jehangir2021}
Jehangir S, Bhat G H, Sheikh J A, Frauendorf S, Li W,  Palit R, and Rather N
2021 \textit{ Eur. Phys. J. A} \textbf{  57} 308
Triaxial projected shell model study of $\gamma$-bands in atomic nuclei

\bibitem{Rajput2022}
Rajput M, Singh S, Verma P, Rani V, Bharti A,  Bhat G H and Sheikh J A
2022 \textit{ Nucl. Phys. A} \textbf{  1019} 122383
Triaxial projected shell model study of $\gamma$-bands in even even $^{104-122}$Cd nuclei

\bibitem{Heenen1993}
Heenen P-H, Bonche P, Dobaczewski J,  Flocard H
1993  \textit{Nucl. Phys. A} \textbf{  561}  367
Generator-coordinate method for triaxial quadrupole dynamics in Sr isotopes (II). Results for particle-number-projected states

\bibitem{Oi2003}
Oi M and Walker P M
2003 \textit{ Phys. Lett. B} \textbf{  576} 75
Three-dimensional rotation of even–even triaxial nuclei

\bibitem{RodriguezGuzman2010}
Rodr\'{\i}guez-Guzm\'{a}n R, Sarriguren P, Robledo L M and Garc\'{\i}a-Ramos J E
2010 \textit{ Phys. Rev. C}   \textbf{   81} 024310 
Mean field study of structural changes in Pt isotopes with the Gogny interaction

\bibitem{RodriguezGuzman2010b}
Rodr\'{\i}guez-Guzm\'{a}n R, Sarriguren P, Robledo L M and Perez-Martin S
2010 \textit{ Phys. Lett. B} \textbf{  691}  202
Charge radii and structural evolution in Sr, Zr, and Mo isotopes

\bibitem{Chen2017}
Chen F-Q and  Egido J L 
2017 \textit{ Phys. Rev. C} \textbf{  95} 024307 
Triaxial shape fluctuations and quasiparticle excitations in heavy nuclei

\bibitem{Niksic2010}
Nik\v{s}i\'{c} T, Ring P, Vretenar D, Tian Y and Ma Z-Y
2010 \textit{ Phys. Rev. C} \textbf{  81} 054318  
3D relativistic Hartree-Bogoliubov model with a separable pairing interaction: Triaxial ground-state shapes

\bibitem{Xiang2016}
Xiang J, Yao J M, Fu Y,  Wang Z H, Li Z P and  Long W H
2016 \textit{ Phys. Rev. C} \textbf{  93} 054324 
Novel triaxial structure in low-lying states of neutron-rich nuclei around $A\approx 100$

\bibitem{Abusara2017a}
Abusara H,  Ahmad S and Othman S
2017 \textit{ Phys. Rev. C}  \textbf{  95} 054302
Triaxiality softness and shape coexistence in Mo and Ru isotopes

\bibitem{Yang2021a}
Yang X Q,  Wang L J, Xiang J,  Wu X Y and Li Z P
2021 \textit{ Phys. Rev. C} \textbf{  103} 054321 
Microscopic analysis of prolate-oblate shape phase transition and shape coexistence in the Er-Pt region

\bibitem{Nomura2021a}
Nomura K, Vretenar D,  Li Z P and Xiang J
2021 \textit{ Phys. Rev. C} \textbf{  103} 054322 
Coupling of pairing and triaxial shape vibrations in collective states of $\gamma$-soft nuclei

\bibitem{Nomura2021b}
Nomura K, Vretenar D, Li Z P and Xiang J
2021 \textit{ Phys. Rev. C} \textbf{  104} 024323 
Interplay between pairing and triaxial shape degrees of freedom in Os and Pt nuclei

\bibitem{ElBassem2024}
El Bassem Y,  El Adri M, El Batoul A and Oulne M
2024 \textit{ Nucl. Phys. A} \textbf{  1043}  122831
Shape evolution and shape coexistence in even-even Mo isotopic chain

\bibitem{Martinou2021a}
Martinou A,  Bonatsos D,  Metzimekis T J, Karakatsanis K E, Assimakis I E,  Peroulis S K, Sarantopoulou S and   Minkov N
2021 \textit{Eur. Phys. J. A} \textbf{57}  84
The islands of shape coexistence within the Elliott and the proxy-SU(3) Models.

\bibitem{Martinou2023}
Martinou  A,  Bonatsos  D,  Peroulis S K,  Karakatsanis  K E,  Mertzimekis T J and  Minkov N
2023 \textit{Symmetry} \textbf{15} 29 
Islands of Shape Coexistence: Theoretical Predictions and Experimental Evidence.

\bibitem{Bonatsos2023}
D. Bonatsos, A. Martinou, S. K. Peroulis, Mertzimekis T J and Minkov N
2023 \textit{ Atoms} \textbf{  11} 117 
Shape Coexistence in Even-Even Nuclei: A Theoretical Overview

\bibitem{Martinou2021b}
Martinou A,  Bonatsos D,  Karakatsanis K E,  Sarantopoulou S,  Assimakis I E,  Peroulis S K  and Minkov  N
2021 \textit{Eur. Phys. J. A} \textbf{57} 83
Why nuclear forces favor the highest weight irreducible representations of the fermionic SU(3) symmetry.

\bibitem{Sarantopoulou2017}
Sarantopoulou S, Bonatsos D, Assimakis I E, Minkov N, Martinou A, Cakirli R B, Casten R F  and Blaum K
2017 \textit{ Bulg. J. Phys.} \textbf{  44}  417
Proxy-SU(3) Symmetry in Heavy Nuclei: Prolate Dominance and Prolate-Oblate Shape Transition

\bibitem{Martinou2020}
 Martinou A, Bonatsos D,  Minkov N,  Assimakis I E,  Peroulis S K,  Sarantopoulou S and  Cseh J
2020 \textit{Eur. Phys. J. A} \textbf{56} 239
Proxy-SU(3) symmetry in the shell model basis

\bibitem{Sorlin2008}
Sorlin O and Porquet M-G
2008 \textit{ Prog. Part. Nucl. Phys.} \textbf{  61} 602 
Nuclear magic numbers: New features far from stability

\bibitem{ensdf}
ENSDF database https://www.nndc.bnl.gov/ensdf

\bibitem{Grosse2022}
Grosse E and  Junghans A R
2022 \textit{ Phys. Lett. B} \textbf{  833}  137328 
Broken axial symmetry as essential feature for a consistent modelling of various observables in heavy nuclei

\bibitem{Bonatsos2020a}
Bonatsos D,  Martinou A,  Sarantopoulou S, Assimakis I E,  Peroulis S and Minkov N
2020 \textit{ Eur. Phys. J. Special Topics} \textbf{229} 2367 
Parameter-free predictions for the collective deformation variables $\beta$ and $\gamma$ within the pseudo-SU(3) scheme.

\bibitem{Casten2000}
Casten R F 2000 \textit{Nuclear Structure from a Simple Perspective}  (Oxford: Oxford University Press)

\bibitem{Otsuka2019}
Otsuka T, Tsunoda Y, Abe T,  Shimizu N and  Van Duppen P
2019 \textit{  Phys. Rev. Lett.} \textbf{  123} 222502  
Underlying Structure of Collective Bands and Self-Organization in Quantum Systems

\bibitem{Egido2020}
 Egido J. L. and  Jungclaus A
2020 \textit{  Phys. Rev. Lett.} \textbf{  125} 192504 
Predominance of Triaxial Shapes in Transitional Super-Heavy Nuclei: Ground-State Deformation and Shape Coexistence along the Flerovium ($Z=114$) Chain of Isotopes

 \bibitem{Zamfir1991}
Zamfir N V and  Casten R F 
1991 \textit{  Phys. Lett. B} \textbf{  260}  265
Signatures of $\gamma$ softness or triaxiality in low energy nuclear spectra

\bibitem{McCutchan2007}
McCutchan E A, Bonatsos D, Zamfir N V and  Casten R F
2007 \textit{  Phys. Rev. C} \textbf{  76} 024306  
Staggering in $\gamma$-band energies and the transition between different structural symmetries in nuclei

\bibitem{Casten2020}
Casten R F, Cakirli R B,  Bonatsos D and  Blaum K
2020 \textit{  Phys. Rev. C} \textbf{  102} 054310 
Simple new signature of structure in deformed nuclei: Distinguishing the nature of axial asymmetry

\bibitem{Bonatsos2021}
Bonatsos D, Assimakis I E,  Martinou A, Sarantopoulou S, Peroulis S K and Minkov N
2021 \textit{  Nucl. Phys. A} \textbf{  1009} 122158
Energy differences of ground state and $\gamma_1$ bands as a hallmark of collective behavior

\bibitem{Draayer1989}
Draayer J P, Leschber Y, Park S C and  Lopez R
1989 \textit{Comput. Phys. Commun.} \textbf{56}  279
Representations of U(3) in U(N)

\bibitem{Pritychenko2016}
Pritychenko B, Birch M, Singh B and Horoi M
2016 \textit{At. Data Nucl. Data Tables} \textbf{107} 1
Tables of E2 transition probabilities from the first $2^+$  states in even–even nuclei


\end{thebibliography}
\end{document}